**Spin-dependent Transport Studies of Fe/$Mo_xCr_{1-x}S_2$/Fe Magnetic Tunnel Junction**

**Aloka Ranjan Sahoo[1,2] and Sharat Chandra[1,2]**

[1]*Materials Science Group, Indira Gandhi Centre for Atomic Research, Kalpakkam, 603102, Tamil Nadu, India*

[2]*Homi Bhabha National Institute, Mumbai, 400094, India*

*Corresponding author's email: alokranjanctc@gmail.com, sharat@igcar.gov.in*

*Abstract*

Using Density functional theory and non-equilibrium Green's function formalism, spin-dependent electron transport in Fe/$Mo_xCr_{1-x}S_2$/Fe magnetic tunnel junction is studied. Spin-transport for different thicknesses (1, 3, 5, and 7 layers) of the spacer $MoS_2$ and for two different surface orientations of the Fe electrode, the Fe(001) and Fe(111) surface and with substitutional doping of 3d transition metal Cr at Mo site in $MoS_2$ is investigated. The electronic structure of the heterostructure shows the presence of metal-induced states in the semiconducting $MoS_2$ at the Fe/$MoS_2$/Fe interface. The I-V characteristics of the junctions for the monolayer and three-layer $MoS_2$ spacer show linear behaviour due to the metallic nature of the junction. The tunnelling nature of the junction is observed for the thicker junctions with five-layer and seven-layer spacers. With the introduction of magnetic impurity Cr, the tunnelling magnetoresistance for 7-layer junction is reduced. The Cr-defect states are observed below the conduction band, and the Cr-doped devices are stable up to a bias of 0.5V. Spin-transport through close-packed Fe(111) surface as electrode show low GMR value.

## 1. INTRODUCTION

A typical spin electronic device, magnetic tunnel junctions (MTJ), is a bi-stable device consisting of a stacked thin, insulating, non-magnetic barrier layer of nm thickness sandwiched between two ferromagnetic electrodes. The resistance of the MTJ can be varied by changing the relative orientation of the magnetisation of the ferromagnetic electrode layers. [1,2] The ferromagnetic electrodes act as a polariser and an analyser of the spin information carried by the current, and the normalised difference between the resistance obtained in these two configuration states is defined as tunnelling magnetoresistance (TMR). The observation of giant magnetoresistance (GMR) in spin-dependent transport in metallic multilayers. [3] The fundamentals of the tunnel magneto-resistive (TMR) and GMR effects enhance memory storage capacity by thousands of times. [4]

Spintronics devices use both the charge and spin degrees of freedom of an electron for the transport and storage of information. Efficient data processing with electron transport and storage of information through an assembly of spins requires the generation, injection and detection of spin-polarised current in a spintronics device. [5–7] Compared to traditional microelectronic devices, spintronics devices function more quickly and generate less heat. Switching a spin state requires less energy than controlling and manipulating a charge current. [8] Potential applications of spintronics include precision sensors, new power-efficient integration architectures, and spin-based fast and ultra-low-power non-volatile devices, such as the new spin-transfer-torque magnetic random-access memory (MRAM) [9,10], widely distributed applications in data storage. [11] This is the demand of Modern Electronics, seeking high-speed data transfer and low-power dissipation. Spintronics aims to provide more complex architectures and a robust solution beyond CMOS. [5,12] Current silicon complementary metal-oxide-semiconductor devices that rely on charge manipulation are most likely to be replaced by devices that manipulate spin.

Traditional MTJ devices prepared with wide band gap oxide spacers such as MgO or $Al_2O_3$ have ill-defined interfaces with ferromagnetic metal electrode in many cases. MTJs with metal oxide-based spacer pose several challenges such as achieving control over layer thickness of the barrier, avoiding pinholes, step interface, and point defects while down to atomic thickness level. [13] Understanding the proximity effects of hybridisation at the interface between ferromagnetic metal and spacer layer is difficult with conventional oxide-based magnetic tunnel junctions due to poor crystallinity and the absence of a smooth interface. In contrast, two-dimensional (2D) materials with atomically thin crystals, free of dangling bonds and control over layer thickness, can be grown on a ferromagnetic electrode layer to provide high-quality, smooth interfaces. [14–16] MTJs with graphene and h-BN as non-magnetic 2D spacer have been demonstrated with well-ordered interface and shows reasonable magnetoresistance value. [17–19] The development of large-scale manufacturing of these 2D materials has further enabled their integration in MTJs. [20–22] Two dimensional (2D) crystals offer a unique potential for spintronic devices due to remarkable properties such as long spin-coherence lengths, [23,24] spin-polarized tunneling, [25] high spin−orbit coupling (SOC) [26] and spin-momentum locking. [27]

The zero band gap and weak spin-orbit coupling (SOC) in graphene make building graphene-based current switches and spintronics applications challenging. In contrast to this, transition metal dichalcogenides are better as barrier layer due to their smaller band gap, resulting lower resistance-area (RA) products. The semiconducting monolayer $MoS_2$ offers quite distinct properties like a high current on-off switching rate ($10^8$), high carrier mobility (~200 $cm^2/Vs$), high thermal stability and ultra-low standby power dissipation have been demonstrated in the field effect transistor [28], tunable band gap, quantum spin Hall effect [29], optical and photo-catalytic properties. [30] In addition, $MoS_2$ and other transition metal dichalcogenides (TMDs) are known to have strong SOC. [31–33] The lack of inversion symmetry coupled with strong SOC leads to The valley polarisation in $MoS_2$ provides a platform for encoding information [34–36] and allows manipulation of the spin and valley degrees of freedom [34,37–40] These 2D materials are very attractive for spintronics, which aims to utilise the spin degree of freedom of electrons for novel information storage and logic devices. [5,7,41,42] The transition metal dichalcogenide (TMDs) $MoS_2$ has a layered structure with Mo sandwiched between two hexagonal planes of S in a trigonal prismatic arrangement. The intra-layer bonding of S-Mo-S is covalent, and inter-layer S-Mo-S sandwiches interact via the weak van der Waals force. [43,44] The indirect band gap of 1.2 eV in bulk $MoS_2$ changes to the direct band gap of 1.8 eV in monolayer $MoS_2$. [45,46] The tunability of the band gap, ultra-thin layered structure, and high carrier mobility make it a suitable candidate for 2D material-based device applications. [47–49]

Oxide-based wide band gap MgO and $Al_2O_3$ serving as a 2D spacer in MTJ has been proposed, and it does not give the desired high magneto resistance (MR) in vertical Spin valves. [50–53] Efficient spin injection and giant magneto resistance (GMR) using $MoS_2$ as a spacer and Fe (001) as a ferromagnetic electrode have been reported by K.Dolui *et al.* [47] They have studied using Fe(001) as an electrode and applied a strain of 10% on Fe to match the lattice of $MoS_2$. Khaldoun *et al.* reported considerable magnetoresistance in a planar Fe/$MoS_2$ junction using a $MoS_2$ nanoribbon. [54] Experimentally, Han-Chun demonstrated a large transverse magnetoresistance at the $Fe_3O_4$/$MoS_2$/$Fe_3O_4$ junction. They observed a clear tunnelling magneto resistance (TMR) signal below 200 K. [48] Recently, Worsak et al. used $Fe_3Si$ Heusler alloy as an electrode for $MoS_2$-based magnetic tunnel junction. [55] Recently, transition metal trichalcogenides-based Weyl half-semimetal with transport along the lateral direction of the 2D materials has been explored for spin-transport. [56] The NiFe/$MoS_2$ interface has also been investigated. The spin-valve effect is observed up to 240 K, with the highest magneto resistance (MR) up to 0.73% at low temperatures, [49] and also from the first-principle electron transport calculations, which reveal MR of ~9% for an ideal Py/$MoS_2$/Py junction. DFT calculation for the electronic structure of interfaces with Fe/$MoS_2$, Co/$MoS_2$, and $Fe_3O_4$/$MoS_2$ has been studied with different surfaces of ferromagnetic material with the $MoS_2$(001) surface. [57]

Interfaces play a crucial role in spin-polarised transport across the ferromagnetic-semiconductor interfaces and are distinct from the individual materials forming the junction. The efficiency of injecting or detecting spin-polarised electrons across such an interface depends on several features, such as interface roughness, defects, impurities, and mismatch in the conductivity of the materials forming the interface. [58–61] The proximity effect of the ferromagnet on the band structure of the semiconducting spacer results in spin-selective transmission. Thus, the tunnelling spin-polarisation is not an intrinsic property of the ferromagnetic electrode alone but depends on the crystalline structure and electronic properties of the insulator and the ferromagnetic/insulator interfaces. [62,63] Understanding the electronic structure of heterostructures is essential in studying their transport properties. The Fe(111) surface has low surface energy, is densely packed, and is the most stable surface. It has a hexagonal symmetry, the same as $MoS_2$, thus creating a low lattice mismatch interface. In the present work, the layer-dependent electronic structure and spin transport in Fe/$MoS_2$/Fe junctions with different ferromagnetic electrode Fe surfaces, Fe(001) and Fe(111), the effect of magnetic impurity Cr-doped $MoS_2$ layers on the spin-transport properties of the junction and the I-V characteristics of the junctions are studied.

## 2. Computational Details

The spin-dependent electron transport studies for Fe/$MoS_2$/Fe interfaces are carried out by combining Non-equilibrium Green's function formalism (NEGF) and the density functional theory (DFT) as implemented in the Quantum-ATK package. [64–66] We have used the local density approximation (LDA) exchange-correlation potential to describe the exchange-correlation effect. [67] The LCAO double-zeta polarised basis set is used for all the elements with a density mesh cut-off of 150 Hartree. [68,69] We have used *bcc* Fe and hexagonal $MoS_2$ with space group $P6_3/mmc$ for bulk calculations and heterostructure construction. The relaxed optimised lattice parameters for Fe and $MoS_2$ with the LDA exchange-correlation functional are 2.77 Å and 3.14 Å, respectively. The Fe(001) plane has ABAB... stacking, and the 2H $MoS_2$(001) has AbA BaB... stacking. The Fe(111) orientation has an ABCDEF... stacking pattern with the spacing along the c-direction of 0.11 Å. The lattice parameters of the Fe(111) electrode are 3.92 Å, 3.92 Å, and 7.20 Å. The Fe/$MoS_2$ interface is made by choosing a 2×1 supercell of Fe(001) and a rectangular unit cell of $MoS_2$ with in-plane lattice parameters of $\sqrt{3}a \times a$. The angle of rotation between the surfaces is 30°, and the angle between the in-plane lattice vectors is 90°, as shown in Figure 1(d). The Fe(001)/$MoS_2$(001) interface is made by choosing a suitable unit cell and applying strain on both Fe and $MoS_2$; the applied strain along the planar direction is 4.88% and -2.3% respectively, with a mean absolute lattice mismatch of 2.39%. [70] The interface geometry's resulting rectangular in-plane lattice parameters are 2.947 Å and 5.495 Å, respectively. The applied strain along the in-plane lattice direction is relaxed by allowing the interface to relax along the *z*-direction. The heterostructure has six, ten and fourteen units of $MoS_2$ in the three-layer, five-layer, and seven-layer device, respectively. For the study of spin transport in Cr-doped $MoS_2$, one Cr atom is substitutionally doped at the Mo site in the middle layer of three-layer, five-layer and seven-layer $MoS_2$ spacers. The Brillouin zone of Fe(001)/$MoS_2$/Fe(001) junction is sampled by a K-point mesh of 13×7×153 using the Monkhorst-Pack scheme. [71] For transmission spectrum calculation, the 2D Brillouin zone is sampled by 75×41 K-points.

For the Fe(111)/$MoS_2$(001) interface, the unit cell for Fe(111) is selected by rotation of the A-lattice vector by 60°, and the new unit cell is formed by the lattice vectors A' and B with in-plane lattice parameter of 2a×2a. For $MoS_2$, a new unit cell is selected by rotation of lattice vectors A by 45° and lattice vector B by 15°, thus the unit cell of $MoS_2$ has A' and B' as lattice vectors with in-plane lattice parameter of $\sqrt{7}a \times \sqrt{7}a$, and a 60° angle between the lattice vectors as shown in Figure 1(e). The relaxed optimised in-plane lattice parameters for Fe(111) and $MoS_2$ with the LDA exchange-correlation functional are 3.92 Å and 3.14 Å, respectively.

The final heterostructure geometry is formed by applying strain on Fe to match the lattice of $MoS_2$. The strain on lattice vector A of Fe is 5.98 % and the final relaxed in-plane lattice parameters of the heterostructure are a = b = 8.318 Å, and the angle between in-plane lattice vectors is 60°. With experimentally grown Fe and $MoS_2$, the Fe(111) and $MoS_2$(001) will have less lattice mismatch of 2% compared to lattice mismatch between Fe(001) and $MoS_2$(001) surfaces. The Brillouin zone is sampled by a Monkhorst-Pack with a grid of 5×5×1, and a tolerance of $10^{-6}$ was set for achieving the self-consistency of electronic energy minimisation. For transmission spectrum analysis, the 2D Brillouin zone is sampled by 51×51 K-points. In all the heterostructure geometries, we have included a sufficient number of Fe layers to use the model for MTJ device calculation to study electron spin-dependent transport through the Fe(001)/$MoS_2$(001)/Fe(001) and Fe(111)/$MoS_2$(001)/Fe(111) sandwich, where it needs to have sufficient number of electrode layers to screen any perturbation due to any scattering by $MoS_2$ to reach the electrode layers. The applied strain on Fe while making the interface is relaxed first by allowing it to relax along the *z*-direction. During relaxing the Fe(111)/$MoS_2$(001)/Fe(111) sandwich, electrode layers are rigidly relaxed as they are bulk-like, and other atoms are entirely relaxed for the target force tolerance of 0.02 eV/Å. In the relaxed geometry, the Fe-S bond distance is found to vary from 2.1Å to 2.29Å, which agrees with the experimental Fe-S bond distance of 2.29 Å. The transition metal Cr has been substitutionally doped at the Mo site in the middle layer for all the devices, and the system is then relaxed as discussed above. The transmission spectrum $T(\varepsilon)$ and spin-dependent current $I_{\uparrow\downarrow}$ are calculated by the formula $T(\varepsilon) = \sum_k t_k t_k^* \delta(\varepsilon - \varepsilon_k)$and $I_{\uparrow\downarrow} = \frac{e}{h}\int_{\mu_L}^{\mu_R} T(\varepsilon)\left[f(\frac{\varepsilon-\mu_L}{k_B T_L}) - f(\frac{\varepsilon-\mu_R}{k_B T_R})\right] d\varepsilon$, respectively. Where $t_k$ is the transmission coefficient from the K-points in the Brillouin zone at energy$\varepsilon$. The chemical potential of the left and right electrodes is denoted by $\mu_L$, $\mu_R$ and $f(\frac{\varepsilon-\mu_L}{k_B T_L})$, $f(\frac{\varepsilon-\mu_R}{k_B T_R})$ are the occupation functions of the left and right electrode respectively at a particular bias voltage. The GMR of the junctions are calculated from the conductance values of parallel ($G_P$) and anti-parallel ($G_{AP}$) configuration by the formula GMR (pessimistic) = $\frac{G_P - G_{AP}}{G_P + G_{AP}} \times 100\%$ and GMR (optimistic) = $\frac{G_P - G_{AP}}{G_{AP}} \times 100\%$.

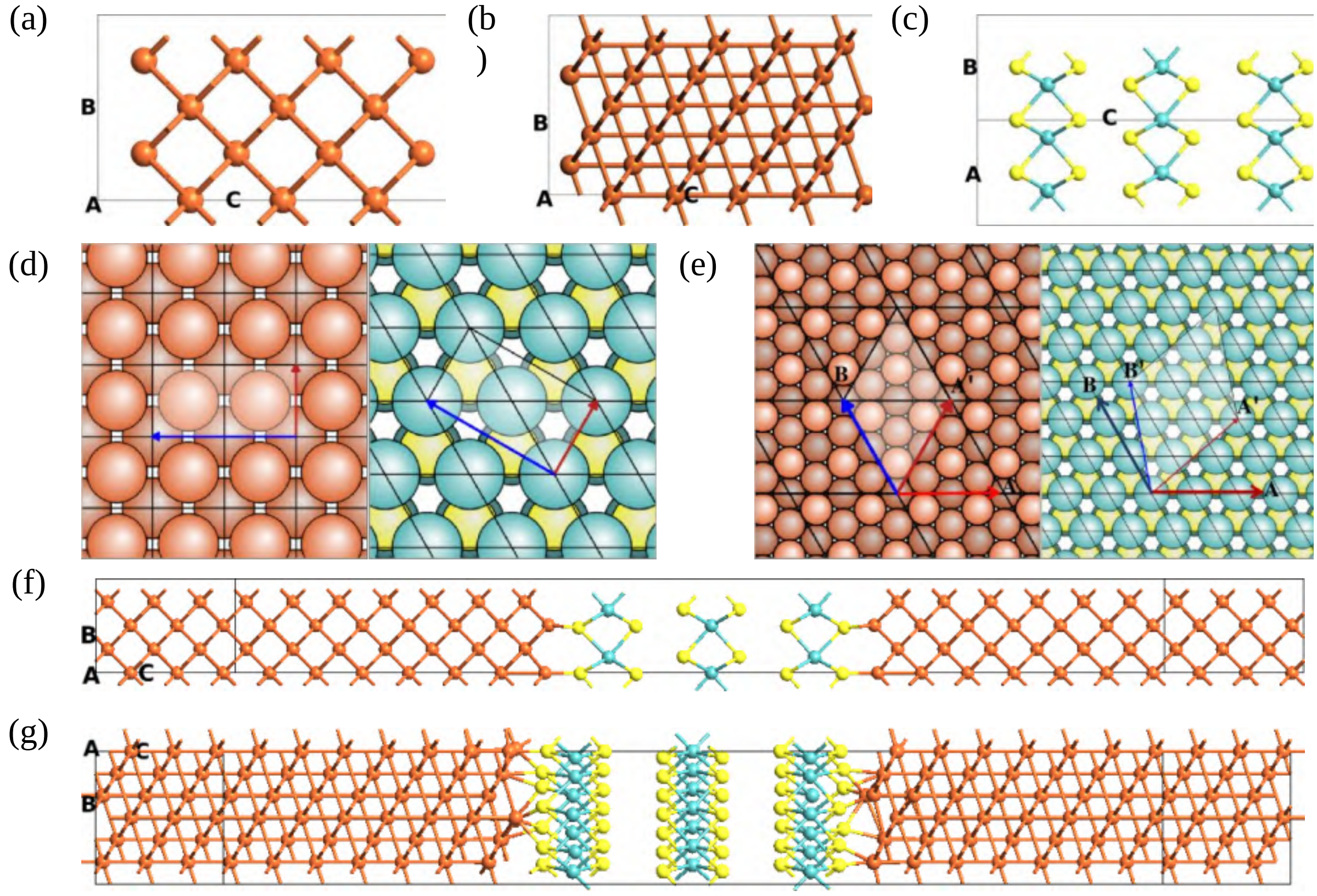

*Figure 1 Stacking pattern of (a) Fe(001), (b) Fe(111) and (c) $MoS_2$(001) surface. Selection of unit cell for heterostructure geometry for Fe(001)/$MoS_2$(001) interface (d) and Fe(111)/$MoS_2$(001) interface (e). Device geometry of Fe/$MoS_2$/Fe Junction with Fe(001) electrode (f) and Fe(111) electrode (g).*

## 3. Results and Discussion

### 3.1. Spin-transport in Fe(001)/$MoS_2$/Fe(001) junction

### **3.1.1.** Electronic structure of interfaces

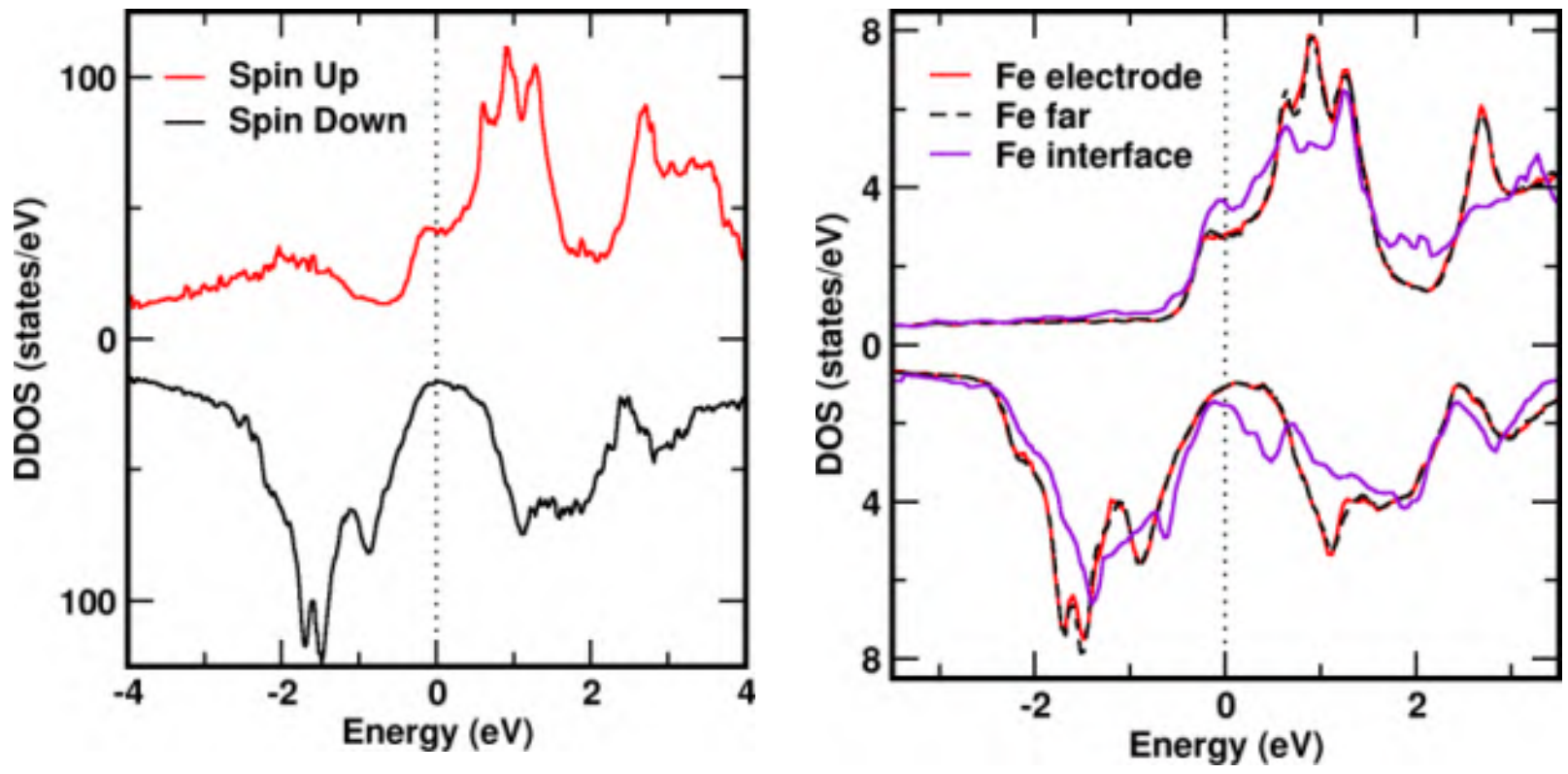

(a) (b)

*Figure 2 Total Device DOS for the seven-layer device (a) and PDOS for Fe layers in the seven-layer device (b).*

The total device density of states (DDOS) and projected density of states (PDOS) for the Fe electrodes, Fe interface layers and Fe layers far away from the interface in the seven-layer device are shown in Figure 2(a) and Figure 2(b), respectively. The PDOS for the Fe layers far from the interface electrode in Figure 2(b) resembles the PDOS for the bulk Fe electrode. This shows that the number of Fe layers is enough to screen the perturbation from the interface. The layer decomposed PDOS for different layers of $MoS_2$ in the central region (in the three, five and seven layer junctions) in Figure 3 shows that the interface layers ($IL_1$ and $IL_2$ $MoS_2$) are always metallic in nature, while the middle layers remain semiconducting. The interface only affects the PDOS of the first two layers of $MoS_2$ ($IL_1$ and $IL_2$ $MoS_2$). Thus, the junction with a three-layer $MoS_2$ spacer is metallic as seen in Figure 3(a) (ML1 $MoS_2$ in the three-layer junction corresponds to $IL_2$ for thicker junctions). A small spin-polarisation is also induced in $MoS_2$ after forming an interface with Fe layers. The PDOS of $MoS_2$ for the junction with five-layer $MoS_2$ as a spacer in Figure 3(b) also shows that while the interface is metallic, the middle layers are semiconducting in nature. The PDOS for the second and fourth interface layers ($IL_2$ $MoS_2$)in Figure 3(b) shows a small spin-up DOS contribution (0.08 states/eV) at the Fermi level, whereas there is a narrow band gap (0.5 eV) in the spin-down channel. The PDOS for the middle layer shows a semiconducting band gap (0.5 eV). Thus, the semiconducting nature of single-layer $MoS_2$ is retained by the middle layer, and it acts as a tunnelling barrier for the electrons. A small spin-polarisation of 10% is seen in the PDOS for the second, third and fourth layers. The layer decomposed PDOS for the seven-layer thick $MoS_2$ spacer in Figure 3(c) shows that the third, fourth and fifth layers (ML1 and ML2 $MoS_2$) of the seven-layer thick spacer maintain the semiconducting nature of $MoS_2$. Spin-polarisation of the DOS for the three layers is also observed. The spin-polarisation contribution arises in the DOS due to the interface with ferromagnetic Fe electrodes. Thus, the seven-layer device shows the largest tunnelling region in the Fe/$MoS_2$/Fe junction.

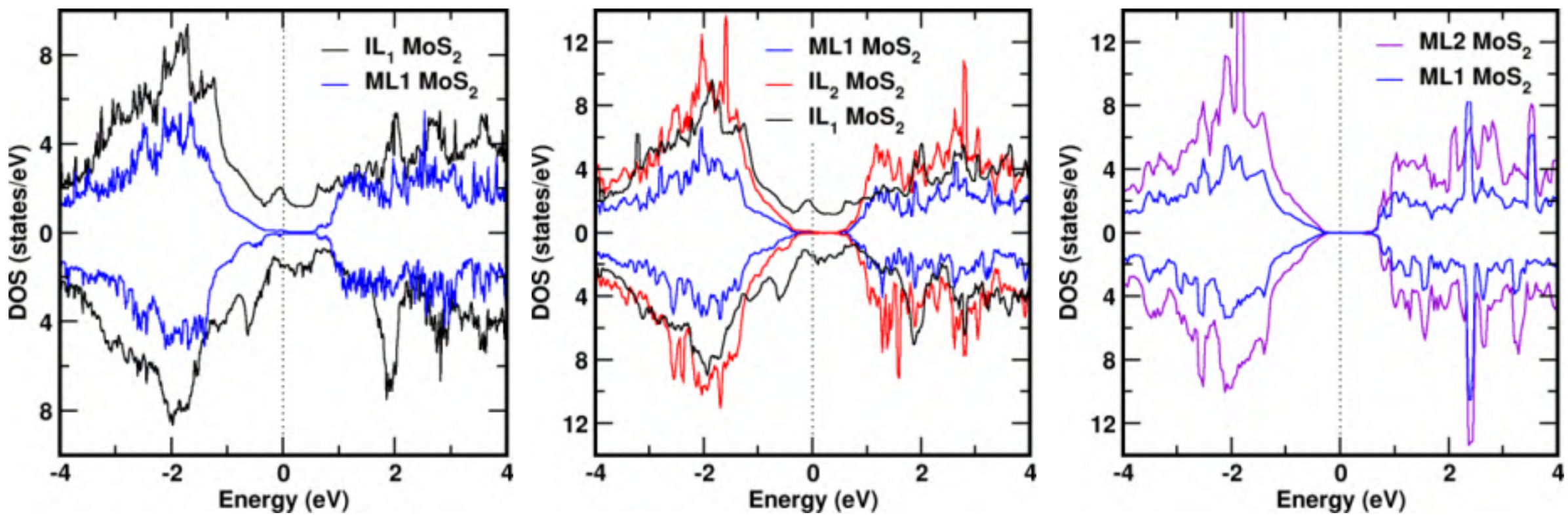

(a) (b) (c)

*Figure 3 Layer decomposed PDOS for $MoS_2$ in three, five and seven-layer Fe(001)/$MoS_2$/Fe(001) junctions in parallel orientation of electrodes.*

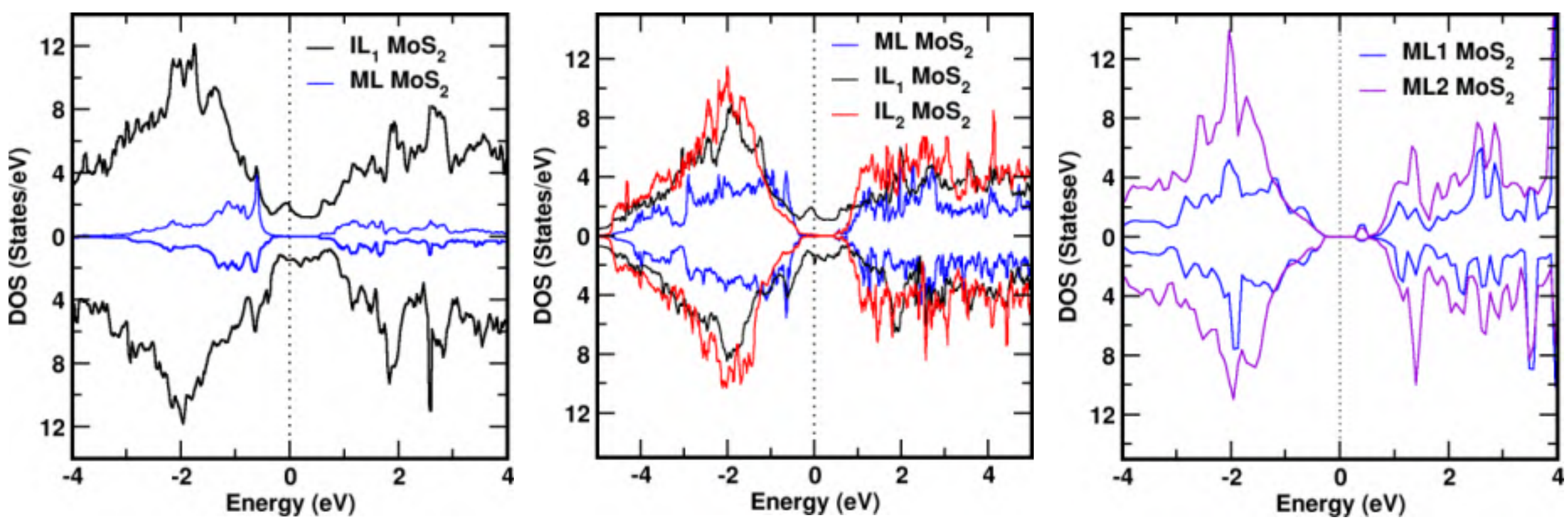

(a) (b) (c)

*Figure 4 Layer decomposed PDOS for $MoS_2$ layers in three, five, and seven-layer Fe(001)/$Mo_xCr_{1-x}S_2$/Fe(001) junction in parallel orientation of electrodes.*

PDOS for the Cr-doped three, five, and seven-layer $MoS_2$ junction devices are shown in Figure 4. PDOS in Figure 4(a) shows that in the trilayer device, the Cr-doped $MoS_2$middle layer (ML $MoS_2$) is metallic, along with the interface layer ($IL_1$ $MoS_2$), after forming a junction with Fe. The middle three layers in the five and seven-layer junctions are semiconducting, as seen in Figure 4(b) and Figure 4(c). The effect of the interface is limited to the first two layers of $MoS_2$ from the interface, as in the undoped system. After doping of Cr does not show much increase in spin-polarisation of the DOS is of 10% .

The projected local density of states (PLDOS) for the undoped and Cr-doped junction with three-layer $MoS_2$ is shown in Figure 5, in which the DOS of the device is projected over the spatial region of the device along the transport direction, which is the *z*-direction. The transport direction *z* is sliced into different regions, and the variation of electronic states with respect to the *z*-direction is shown in the PLDOS. The spatially resolved density of states is shown in the PLDOS with a 2D colour map showing the distribution of electronic states over both energy and real space. Scanning across the *z*-direction shows the change in local electronic structure across the electrode, interface and central scattering region. Brighter red dots show a high presence of states, and a dark/blue dot shows a low presence of states in the energy range and the atomic layer contributing to the DOS. A dark black region in the middle of the figures shows an absence of states, indicating a band gap for the semiconductor or insulator spacer in the device.

From the 2D colour map for the PLDOS, light green regions can be seen on both sides of the Fe/$MoS_2$ interface corresponding to the first layer of $MoS_2$ at the interface, showing the presence of states in the band gap region of $MoS_2$. The strong coupling between the ferromagnetic Fe layers and the first layer of $MoS_2$ at the Fe/$MoS_2$ interface is observed, and the coupling is strong enough for the next $MoS_2$ layer to exhibit metal-induced states on both sides of the interface. The metallicity of the interface $MoS_2$ arises from the strong hybridisation of Fe *d*-orbitals with *d*-orbitals of Mo and *p*-orbitals of S. For the next layer of $MoS_2$ at the interface, only the spin-up channel crosses the Fermi-level, as seen in the PDOS for the three-layer $MoS_2$ in Figure 3(b). Metal-induced states are present in all three layers of $MoS_2$, with the first layer of $MoS_2$ next to the interface Fe layers having higher metal-induced states.

The Cr doped middle layer of $MoS_2$ spacer in the three-layer $Mo_xCr_{1-x}S_2$is metallic, due to the interaction with the Fe interface. These can be seen in the PLDOS for both the spin-up and spin-down channels in Figure 5(c) and (d). The PLDOS for the middle layer shows that the presence of states in the conduction band has shifted to lower energies after the substitutional doping of Cr in $MoS_2$.

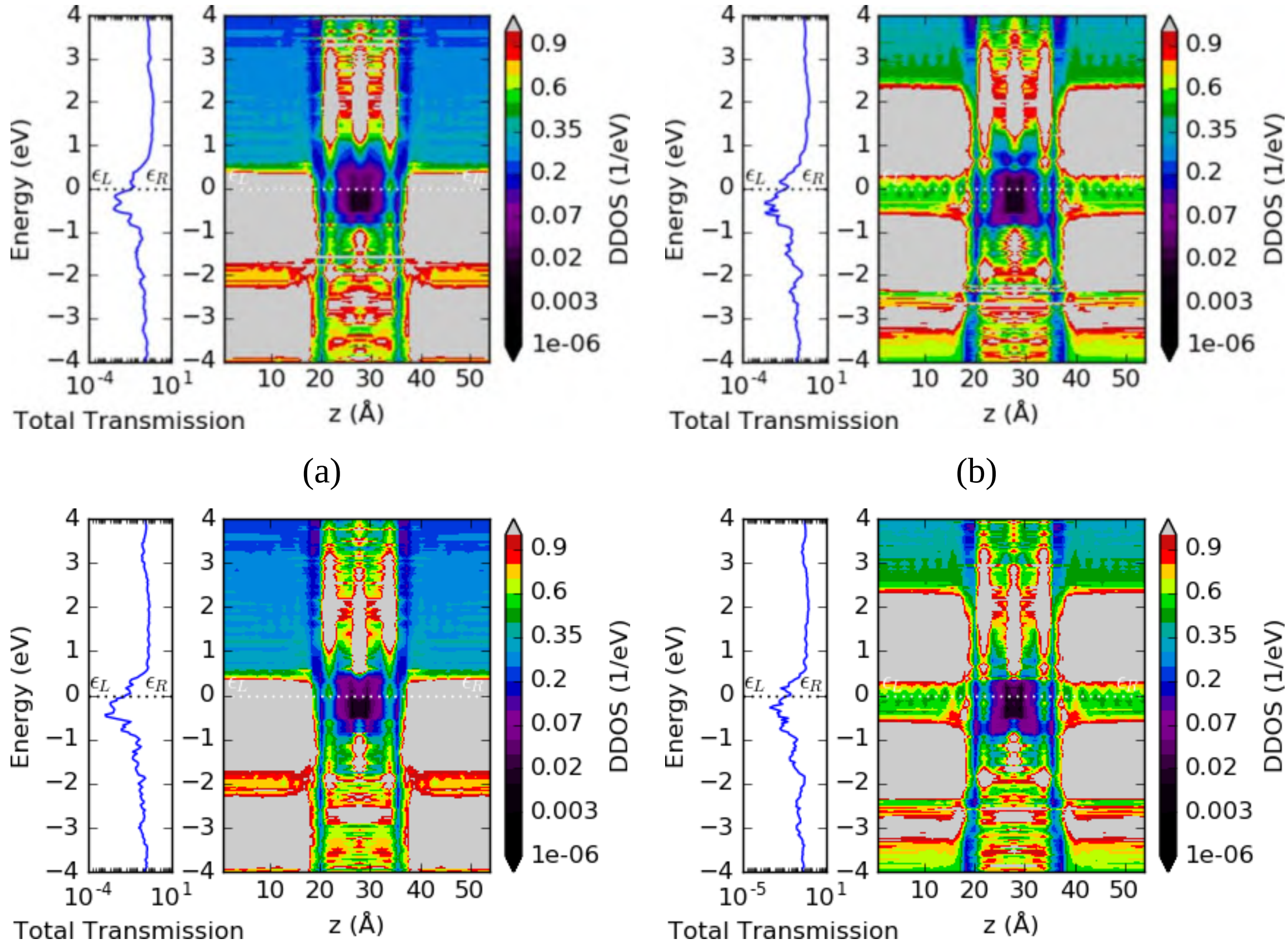

*Figure 5 PLDOS for undoped (upper panel) and Cr-doped (lower panel) Fe(001)/$Mo_xCr_{1-x}S_2$(3-layer)/Fe(001) junction in the spin-up and spin-down channels in parallel orientation of electrodes.*

The PLDOS for five-layer and seven-layer thick $MoS_2$ as a barrier in the Fe/$MoS_2$/Fe junction and Cr-doped junctions in parallel magnetic orientation is given in Figures S1 to S4 of the supplementary data. The middle layer of $MoS_2$ in the five-layer, and the subsequent three layers of $MoS_2$ next to the interface layer in the middle of the seven-layer-thick $MoS_2$ retain the semiconducting nature of the band gap, as seen in the PLDOS and PDOS for the middle layer in Figure 3(b) and Figure 3(c), respectively. Thus, the effect of coupling is limited to the first two layers of $MoS_2$ at the Fe/$MoS_2$ interface. The PLDOS of the Fe/$Mo_xCr_{1-x}S_2$(5-layer)/Fe and Fe/$Mo_xCr_{1-x}S_2$(7-layer)/Fe junction in parallel orientation shows similar behaviour for the five-layer and seven-layer devices. After doping with Cr, the PLDOS shows that the conduction band for the middle layer moves down in energy, which can be seen from the bright, lighter regions for the PLDOS of the middle layer in Figure 5(c) and (d). The presence of states in the band gap region of the Cr-doped middle layer $MoS_2$is due to the presence of Cr defect levels. Thus, the band gap for the Cr-doped $MoS_2$ reduces due to the presence of Cr-defect levels in the band gap region, as observed in previous studies. [72,73]

### **3.1.2.** Spin transport in Fe(001)/$MoS_2$/Fe(001) Junction

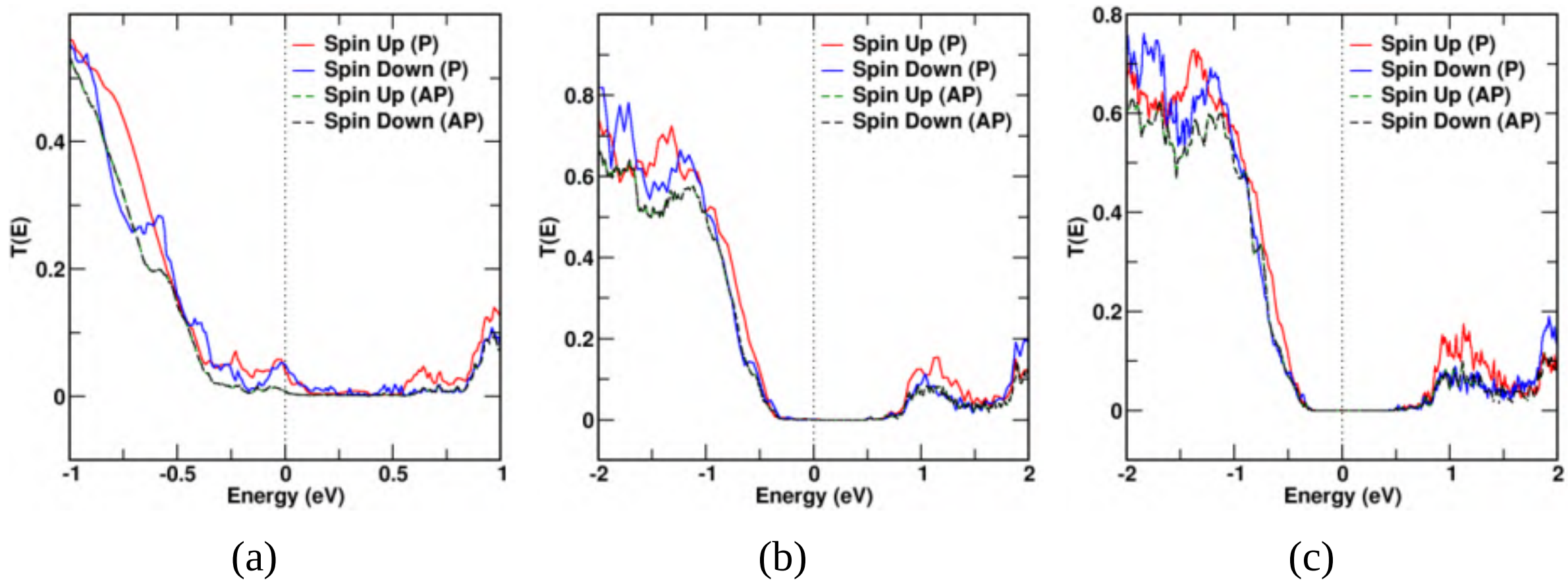

*Figure 6 Transmission spectrum for Fe(001)/$MoS_2$/Fe(001) junction in parallel (P) and anti-parallel (AP) orientations of electrodes for three-layer (a), five-layer (b), seven-layer (c) thick $MoS_2$ spacer.*

The metallic nature of the three-layer thick junction leads to the finite transmission through the junction under zero-bias conditions. The transmission spectrum for the 3, 5 and 7-layer devices in parallel orientation of the electrodes (P) is shown in Figure 6. For the three-layer device, the transmission at the Fermi level is 0.04 and 0.05 in the spin-up and spin-down channels, respectively. For the anti-parallel orientation of the electrodes (AP), the junction has 0.008 transmissions for both spin-up and spin-down channels at the Fermi level, as seen in Figure 6(a). For the Fe/$MoS_2$(5-layer)/Fe junction, the transmission for the device in P orientation has significantly reduced at zero-bias with values 0.0004 and 0.0002 for the spin-up and spin-down channels, respectively. For AP orientation of electrodes, the transmission for spin-up and spin-down channels is the same and of the order of $4\times10^{-5}$, as seen in Figure 6(b). The transmission spectrum for the junction with seven-layer $MoS_2$ as a spacer at zero-bias, the transmission spectrum for the junction across the spin-up and spin-down channels is $1.6\times10^{-5}$ and $1.75\times10^{-6}$, respectively and for the AP orientation of the electrodes, the transmission spectrum is the same for both channels. It has a transmission of the order of $10^{-6}$ for both spin-up and spin-down channels, as shown in Figure 6(c).

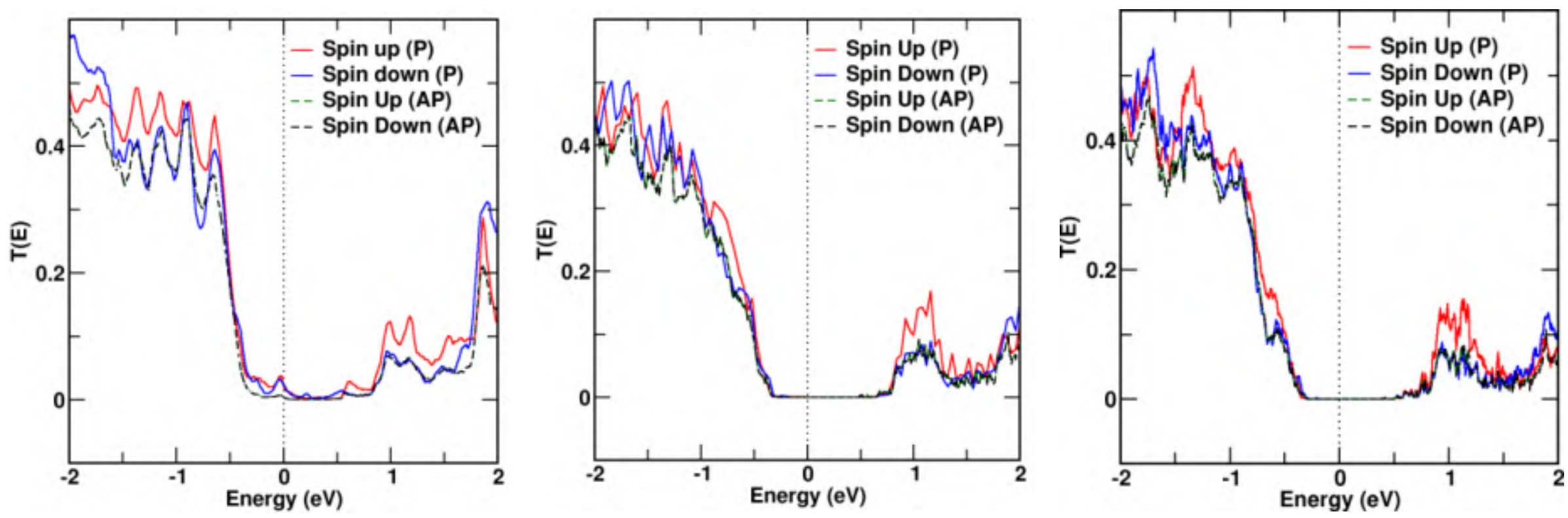

(a) (b) (c)

*Figure 7 Transmission spectrum for Cr-doped three-layer (a), five-layer (b), seven-layer (c) Fe(001)/$Mo_xCr_{1-x}S_2$/Fe(001) junction.*

With Cr doped in the middle layer of the three-layer $MoS_2$ in Fe/$Mo_xCr_{1-x}S_2$(3-layer)/Fe junction, the transmission for both spin-up and spin-down channels in the parallel orientation of the electrodes is 0.025. In the anti-parallel orientation of the electrodes, the transmission is 0.004 for each channel, as shown in Figure 7(a). For the Cr-doped five-layer $MoS_2$ spacer, the spin-dependent transmission for the Fe/$MoS_2$(5-layer)/Fe device at zero-bias condition is shown in Figure 7(b). The P orientation has a transmission of $1.5\times10^{-4}$ in spin-up, and spin-down has the order of $10^{-5}$ at the Fermi-level. For the AP orientation of electrodes, the junction has transmission of $3.5\times10^{-5}$ for both spin-up and spin-down channels. The transmission spectrum for seven layers of thick $MoS_2$ doped with Cr in the middle layer of $MoS_2$, the transmission spectrum for both P and AP orientation of electrodes at the Fermi level is of the order of $10^{-6}$ and for AP, the transmission spectrum is identical for both spin-up and spin-down channels, as shown in Figure 7(c).

The K-resolved transmission spectrum for P is shown in Figure 8. The contribution from the 2D Brillouin zone to the transmission spectrum at each energy is shown with a 2D colour map. The brighter red dots show a higher transmission coefficient, indicating a higher contribution from the K-point at a particular energy, and a dark region shows a low transmission coefficient from the K-point at a specific energy, indicating a low contribution from the K-point.

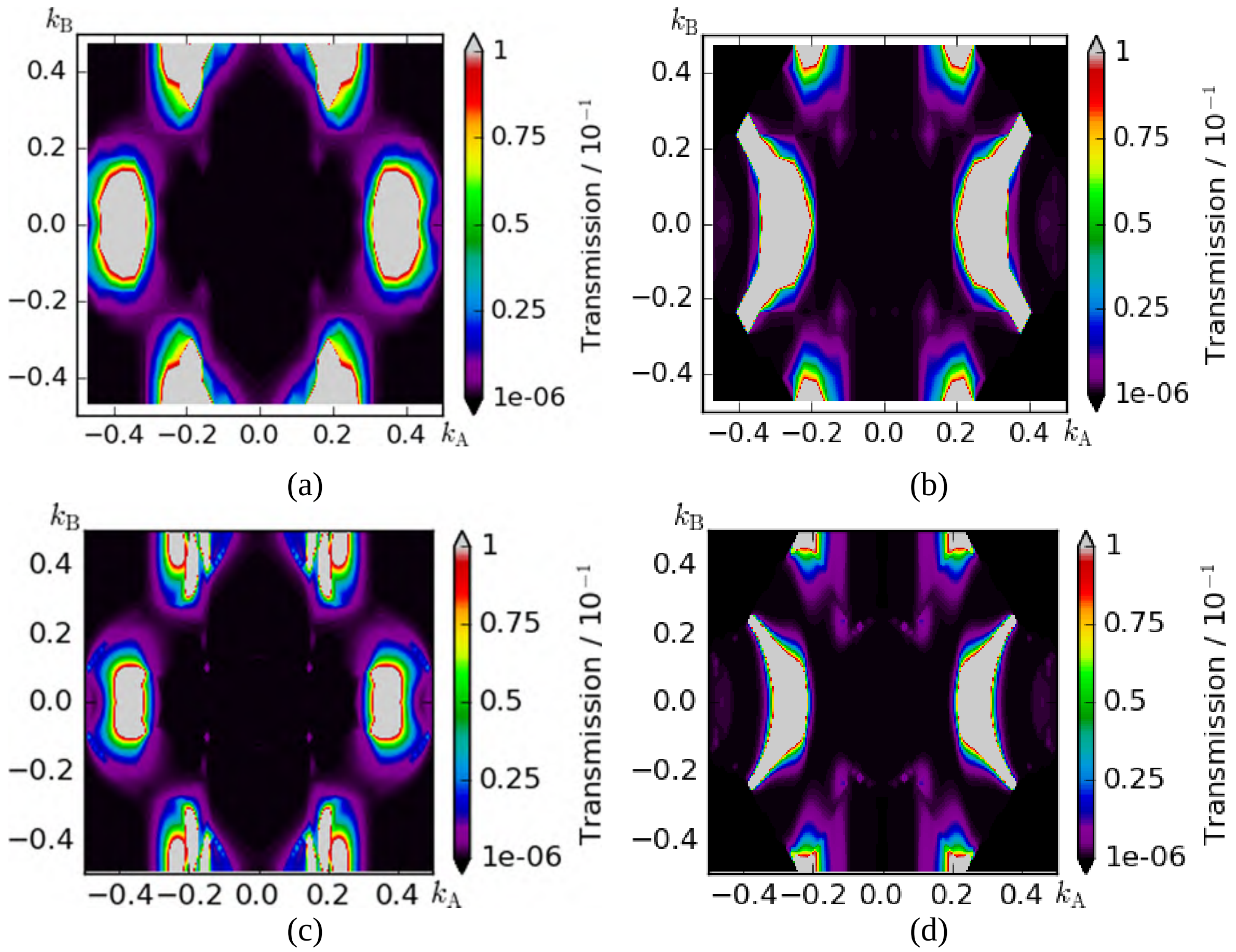

*Figure 8 K-resolved transmission spectrum for undoped (figure a, and b) and doped (figure c, and d) Fe(001)/$Mo_xCr_{1-x}S_2$(3-layer)/Fe(001) junction in parallel orientation of electrodes in spin-up (left panel) and spin-down (right panel) channel.*

The K-resolved transmission spectrum for parallel orientation of the electrodes under zero-bias shows that the contribution to the transmission at the Fermi-level comes from six pockets away from the 2D Brillouin zone centre in the spin-up channel, and for spin-down, two pockets majorly contribute to the transmission, as seen in Figure 8. For the anti-parallel orientation of the electrodes, the K-resolved transmission for both the spin channels is identical. It comes from the other four pockets of the Brillouin zone (refer to Figure S5 of the supplementary data).

The K-resolved transmission spectrum for 5-layer and 7-layer thick undoped and Cr-doped junctions (see Figures S6 to S8 in the supplementary data) shows that the contribution to the total transmission comes from six points and four points from the 2D Brillouin zone in P and AP, respectively. With an increase in layer thickness of the spacer, the transmission coefficient decreases from $1\times10^{-2}$ in the 5-layer thick junction to $10^{-4}$ in the 7-layer thick junction in the spin-up channel and from $10^{-3}$ to $10^{-5}$ in the spin-down channel.

### **3.1.3.** GMR and I-V characteristics of the junctions

| Configuration $Fe(001)/MoS_2/Fe(001)$ | | Parallel Orientation | | Anti-Parallel Orientation | | TMR | |
|---|---|---|---|---|---|---|---|
| | | ↑ spin (Siemens) | ↓ spin (Siemens) | ↑ spin (Siemens) | ↓ spin (Siemens) | Optimistic (%) | Pessimistic (%) |
| 3 Layers | Undoped | $1.35\times10^{-6}$ | $1.80\times10^{-6}$ | $2.60\times10^{-7}$ | $2.68\times10^{-7}$ | 496.74 | 71.30 |
| | Cr doped | $7.96\times10^{-7}$ | $9.33\times10^{-7}$ | $1.59\times10^{-7}$ | $1.60\times10^{-7}$ | 440.97 | 68.80 |
| 5 Layers | Undoped | $1.85\times10^{-8}$ | $7.15\times10^{-9}$ | $1.84\times10^{-9}$ | $1.81\times10^{-9}$ | 602.71 | 75.08 |
| | Cr doped | $1.60\times10^{-8}$ | $2.30\times10^{-9}$ | $1.74\times10^{-9}$ | $1.73\times10^{-9}$ | 426.55 | 74.13 |
| 7 Layers | Undoped | $1.82\times10^{-9}$ | $1.27\times10^{-10}$ | $1.45\times10^{-10}$ | $1.45\times10^{-10}$ | 573.05 | 74.13 |
| | Cr doped | $5.00\times10^{-10}$ | $4.87\times10^{-11}$ | $8.36\times10^{-11}$ | $8.27\times10^{-11}$ | 229.65 | 53.45 |

*Table 1 GMR measurements for undoped and Cr-doped $Fe(001)/MoS_2/Fe(001)$ junctions.*

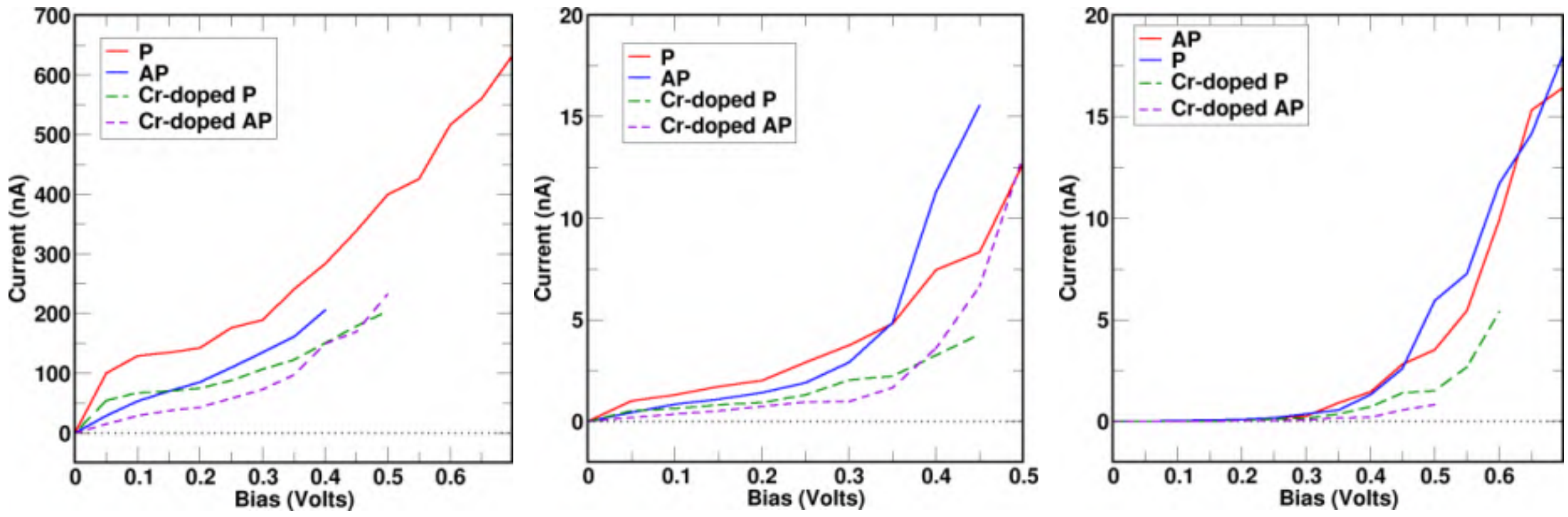

(a) (b) (c)

*Figure 9 I-V plot for undoped and doped devices in 3-layer (a), 5-layer (b)and 7-layer (c).*

The I-V plot for the 3-layer device with Cr-doped $MoS_2$ and undoped $MoS_2$ as a spacer is shown in Figure 9(a). The I-V plot for the 3-layer device shows linear behaviour due to the metallic nature of the first $MoS_2$ layer and the half-metallic nature of the next layer at the interface. The device is stable up to a bias of 0.5V, and the spin current for the parallel configuration is higher than that of the anti-parallel configuration. For 5-layer $MoS_2$ as a spacer, the spin-current has further reduced and is of the order of 10 nano-Ampere, as the interface effect is limited only to the first two layers of $MoS_2$. The tunnelling behaviour of electrons in the junction is observed. After a bias of 0.35V, the spin current for the anti-parallel configuration is higher than that of the parallel configuration. This is also seen in the K-resolved transmission spectrum for the device, where the transmission coefficient for the spin-down channel is higher than the spin-up channel. The 5-layer device is stable up to a bias of 0.5V. The bias range for the device is understandable because the band of $MoS_2$ is of the order of 0.7V with LDA approximation, where we see that with Cr doping in the $MoS_2$ layer, the presence of Cr-defect levels is below the conduction band. I-V characteristics for the junction with 7-layer $MoS_2$ as the spacer are shown in Figure 9(c). The device is stable up to a bias of 0.7V and 0.5V for undoped $MoS_2$ and Cr-doped $MoS_2$, respectively. The tunnelling nature of the junction is observed as the middle three layers of the 7-layer $MoS_2$ keep the semiconducting nature of the bulk $MoS_2$. GMR for the junction is 74 %, and with Cr-doping, the device's GMR is reduced. The Cr-doped devices are stable up to 0.5 V bias and the undoped devices are stable up to 0.7 V bias. The stability of the device with respect to the bias voltage depends on the electronic structure of the device. With LDA exchange-correlation functional, the undoped $MoS_2$ monolayer has a direct band gap of 1.88 eV and bulk 2H-$MoS_2$ has direct band gap of 0.5V and indirect bandgap of 1.757 eV. With increase in layer thickness of 2H-$MoS_2$ the band gap decreases and approaches the bulk value in multi-layers. Therefore the 7-layer $MoS_2$ has a lower indirect band gap in comparison to 5-layer and 3-layer $MoS_2$. The monolayer has the largest band gap followed by 3-layer, 5-layer and 7-layer $MoS_2$. With substitutional Cr-doping at the Mo site the band gap of 2H-$MoS_2$ decreases with appearance of Cr-defect levels below the conduction band of undoped $MoS_2$. Therefore, when the bias voltage is increased beyond the electronic band gap of the spacer, more channels become available for conduction of electrons thereby causing the short circuit and the junction is destabilized.

### **3.1.4.** Effects of Cr doping in $MoS_2$

Effects of Cr doping in $MoS_2$ have been studied by many groups. [72,74–77] The Cr doping concentration varies from 5% to 15%. Changes in the crystalline structure, band gap, magnetic moments, supercapacitance and the electrical characteristics of the FET devices have been studied in these works. Cr is shown to substitutionally dope Mo [74]. Mo site in $MoS_2$ is preferred for Cr doping of 3 to 7% in $MoS_2$ as observed from the XRD, XPS and AFM measurements. Higher doping concentration of 15% leads to reduction in crystallinity and Cr clustering is observed as the doping concentration increases [75]. The band gap decreases upon doping in monolayer $MoS_2$ [72]. The remnant magnetization and coercivity decrease at low temperatures indicating a reduction in the magnetic interactions [76], supercapacitance decreases [74], and while the electrical mobility decreases [72], the carrier concentration increases resulting in faster switching, on/off ratio in FET transistors. It is shown in [77] that the undoped samples shows higher MR value of 110% and with doping of 5% and 8% Cr-doped samples have 2H and 1T mixed phases. They show negative MR of -8% and -12% at 100 Hz frequency. For higher doping concentration of 15% the sample has Cr clustering and the MR turns positive and is 60%.

We have carried out calculations for 7% to 17% Cr doping in $MoS_2$ multilayers which correspond to the doping concentration across the range of the published experimental results. There is no Cr clustering in our calculations and while the MR remains similar to that of the undoped case for 3-layer and 5-layer $MoS_2$ (Cr doping concentration of 17% and 10%), it decreases by 28% for the 7-layer (Cr doping concentration 7%) case. The indirect band gap is seen to decrease with increase in layer thickness and as a function of Cr doping and there is not much change seen in the spin polarization ratio. So, our results largely confirm the behavior seen in the experiments.

### **3.2.** Spin transport in Fe(111)/$MoS_2$/Fe(111) Junction

#### **3.2.1.** Electronic structure of Fe(111)/ $MoS_2$/Fe(111)junction

Figure 10 shows the transmission spectrum(upper panel) for the one, three and five layer thick $MoS_2$ spacer in the Fe(111)/$MoS_2$/Fe(111) junction, and Figure 10( lower panel) shows PDOS for single-layer, three-layer and five-layer $MoS_2$ in the junction. PDOS for the $MoS_2$ shows the presence of states at the Fermi-level for the single-layer $MoS_2$ as a spacer. For the three-layer $MoS_2$, the interface $MoS_2$layer is also metallic, and the PDOS for the middle layer of the three-layer $MoS_2$ shows that the DOS for spin-up and spin-down electrons touches the Fermi level, as shown in Figure 10(b). For a thicker junction of five layers of $MoS_2$, the PDOS for the middle layer of $MoS_2$ in the five-layer $MoS_2$ is shown in Figure 10(c). From the PDOS, it's evident that the thickness of five layers of $MoS_2$ as the spacer is thick enough to keep the semiconductor behaviour of the middle layer of $MoS_2$ and to reduce the effect of the interface.

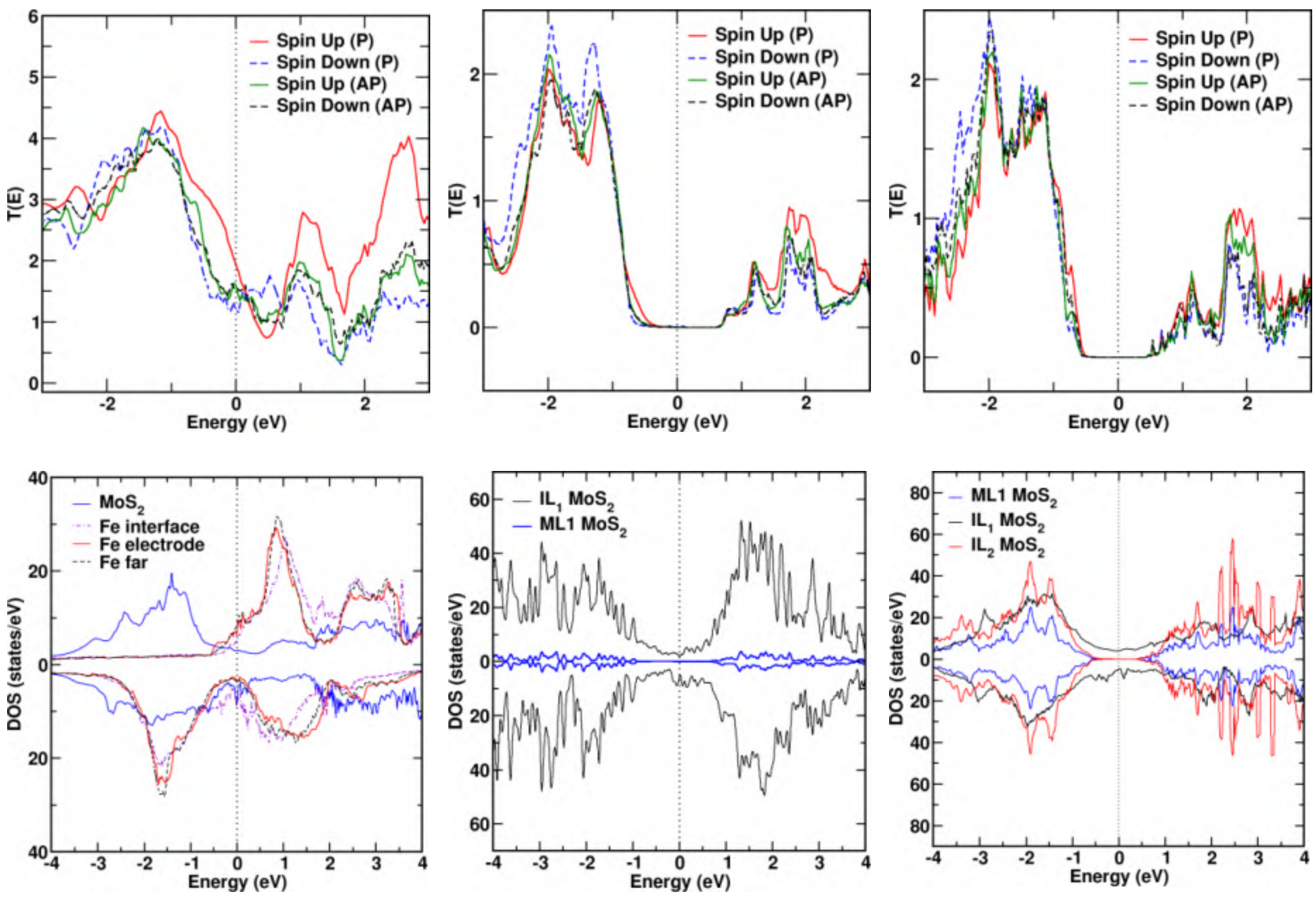

(a) (b) (c)

*Figure 10 Transmission spectrum (upper panel) and PDOS (lower panel) of single-layer $MoS_2$ (a), 3-layer $MoS_2$ (b), and 5-layer $MoS_2$(c) in the Fe(111)/$MoS_2$/Fe(111) junctions.*

The transmission spectrum for P and AP of the electrodes in the Fe(111)/$MoS_2$(1,3,5 layer)/Fe(111) junction under zero-bias conditions is shown in Figure 10 (upper panel). For the parallel orientation of the electrode, the transmission for spin-up and spin-down channels at the Fermi level is 1.89 and 1.26, respectively. The metallic nature of the junction leads to higher transmission at zero bias. Under the AP orientation of the electrodes, the transmission across the spin-up and spin-down channels is identical; the transmission at the Fermi level for the channel is 1.63, as shown in Figure 10(a). For the 3-layer junction, the P orientation of electrodes under zero bias, the transmission at the Fermi level is $1.2\times10^{-3}$ and $4\times10^{-3}$ for spin-up and spin-down channels, respectively. For the AP orientation of electrodes, the transmission for both spin-up and spin-down channels is $2.0\times10^{-3}$ and $2.6\times10^{-3}$, respectively, as seen in Figure 10(b). The transmission spectrum for the Fe(111)/$MoS_2$(5-layer)/Fe(111) junction in P and AP orientation of electrodes under zero-bias conditions is shown Figure 10(c) shows for the P, the transmission at the Fermi level is of the order of $3\times10^{-6}$ for both spin-up and spin-down channel and for AP, the transmission spectrum at the Femi-level for both channels are the same and is of the order of $3.4\times10^{-6}$.

The PLDOS for the one, three and five-layer Fe(111)/$MoS_2$(1-layer)/Fe(111) junction is shown in Figure 11. The 2D colour map for the PLDOS shows the effect of interface Fe layers on the LDOS of $MoS_2$in both spin-up and spin-down channels. With the presence of metal-induced states in both channels of the single-layer $MoS_2$ spacer, the number of states available at the Fermi level of $MoS_2$ is higher for the spin-down channel than for the spin-up channel. For the 3-layer junction, metal-induced states are present in the first layer of $MoS_2$ on both sides of the interface and the middle layer $MoS_2$ shows DOS for spin-up and spin-down touching the Fermi level. For the Fe(111)/$MoS_2$(5-layer)/Fe(111) junction, the metal-induced states are present for the first layer of $MoS_2$ on both sides of the interface, and the middle layer of $MoS_2$ keeps the semiconducting band gap nature of bulk $MoS_2$ (refer to Figure S9 in supplementary data).

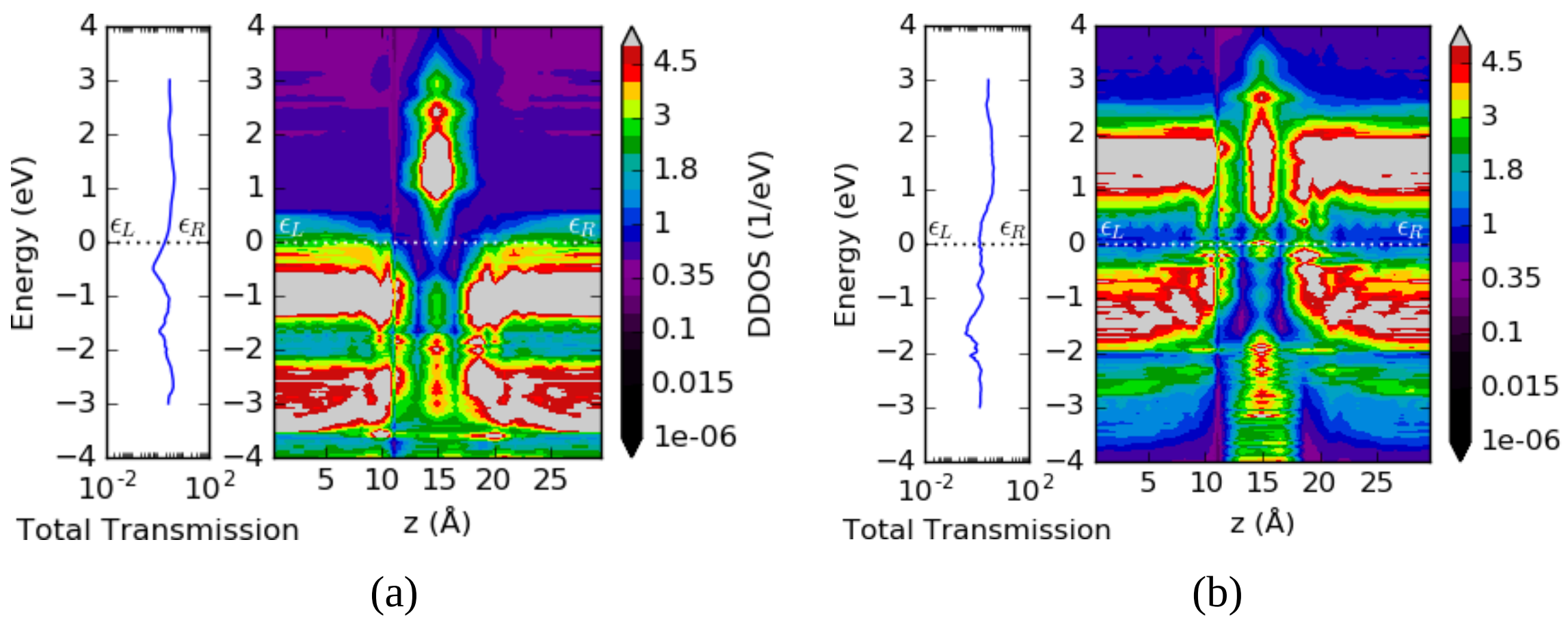

*Figure 11 PLDOS of one-layer Fe(111)/$MoS_2$/Fe(111) junction in the parallel orientation of electrodes for the (a) spin-up and (b) spin-down channel.*

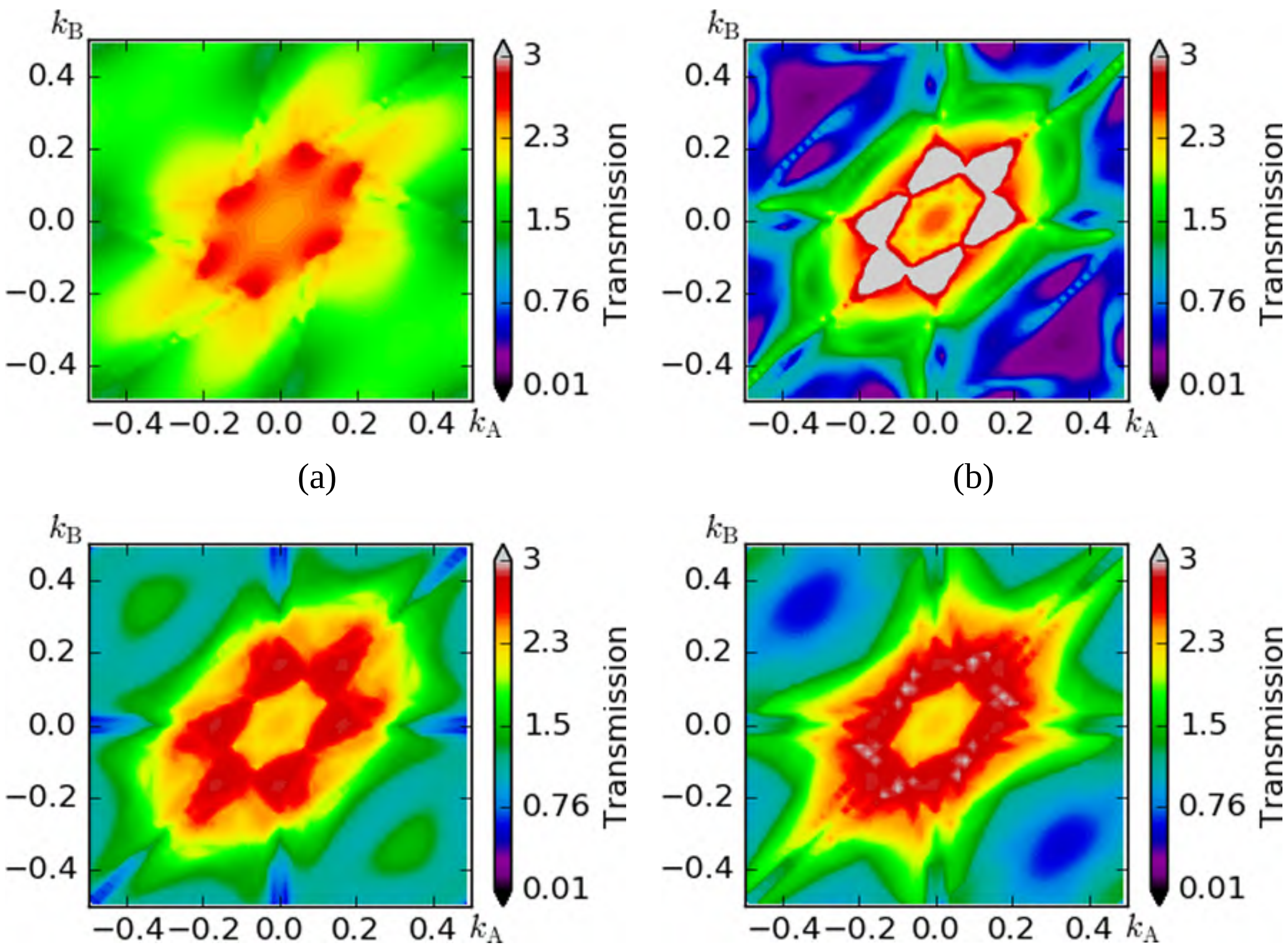

*Figure 12 K-resolved transmission spectrum for spin-up (a) and spin-down (b) channel in Fe(111)/$MoS_2$(1-layer)/Fe(111) junction in the parallel (a) and (b) and anti-parallel (c) and (d) orientation of electrodes.*

The K-resolved transmission spectrum for the Fe(111)/$MoS_2$(1-layer)/Fe(111) junction in P and AP is shown in Figure 12. In the parallel orientation of the electrodes, the spin-up channel gets a contribution from the entire 2D Brillouin zone, having a transmission coefficient of 1 from all the parts, and the major contribution for a value close to 3 comes from the six corners close to the centre of the 2D hexagonal Brillouin zone. For the spin-down channel, the transmission coefficient has a maximum value of 3 from the six hexagonal arms of the hexagonal 2D Brillouin zone near the Brillouin zone centre. For the anti-parallel orientation of electrodes, the K-resolved transmission is a combination of both channels of the P case, with a maximum value of 3 coming from the regions close to the Brillouin zone centre.

For devices with three- and five-layer $MoS_2$ spacer, the transmission in P at the Fermi-level comes from six points in the 2D Brillouin zone with the transmission coefficient of the order $10^{-3}$ and $10^{-2}$ in the spin-up and spin-down channels, respectively. For AP, it's $5\times10^{-3}$ for both the spin channels. The 2D color map shows the transmission in the spin-up and spin-down channels for the AP, which is a combination of both channels in P. The transmission for the spin-down is higher due to the DOS in the spin-down channel crossing the Fermi level. With an increase in layer thickness of the spacer, the transmission coefficient decreases by an order of $10^3$ for the five-layer $MoS_2$ spacer with transmission coefficients of $10^{-6}$ and $10^{-5}$ along the spin-up and spin-down channels, respectively (refer to Figure S10 and S11 in supplementary data).

| Configuration Fe(111)/$MoS_2$/Fe(111) | | Parallel Orientation | | Anti-Parallel Orientation | | GMR | |
|---|---|---|---|---|---|---|---|
| | | ↑ spin (Siemens) | ↓ spin (Siemens) | ↑ spin (Siemens) | ↓ spin (Siemens) | Optimistic (%) | Pessimistic (%) |
| 1 Layers | Undoped | $7.90\times10^{-05}$ | $5.52\times10^{-05}$ | $6.87\times10^{-05}$ | $6.83\times10^{-05}$ | -2.0% | -1.0% |
| 3 Layers | Undoped | $5.25\times10^{-08}$ | $2.18\times10^{-07}$ | $9.88^{-08}$ | $1.10\times10^{-07}$ | 29.98 | 13.03 |
| 5 Layers | Undoped | $9.46\times10^{-11}$ | $1.24\times10^{-10}$ | $1.18\times10^{-10}$ | $9.41\times10^{-11}$ | 3.17 | 1.56 |

*Table 2 GMR measurement for Fe(111)/$MoS_2$/Fe(111) junction.*

GMR measurement for the one, three, and five-layer device shows that the GMR values for the junctions Fe(111) electrode yield lower GMR values than the device with Fe(001) electrode due to the availability of channels in spin-down DOS.

## 4. Conclusion

Spin-dependent transport studies through Fe(001)/$Mo_xCr_{1-x}S_2$/Fe(001) junction with a thickness of three, five and seven layers of $MoS_2$ as spacer material and spin transport across the low-strain Fe(111)/$MoS_2$/Fe(111) junction with a thickness of one, three and five layers of $MoS_2$ is studied. The Fe(111)/$MoS_2$/Fe(111) junctions show low GMR value. The Fe/$MoS_2$ interface with one layer and three layers of $Mo_xCr_{1-x}S_2$ as the spacer is metallic. The metallic nature of the interface comes from the strong coupling of Fe layers with $MoS_2$ layers at the interface. The strong hybridisation of Fe *d*-orbitals with *p*-orbitals of S and *d*-orbitals of Mo induces metal-induced states in the semiconducting $MoS_2$. The hybridisation is strong for the first layer of $MoS_2$ at the interface. For the next layer of $MoS_2$ (the second layer next to the interface), the PDOS shows only one of the spin channels crossing the Fermi level, and the other has a band gap region. Thus, the second layer of $MoS_2$ near the interface with Fe behaves like a nearly half-metallic character. At zero bias, the transport through a single layer of $MoS_2$ and Fe(111) as the electrode is high for the P orientation of the electrodes. For the AP orientation, the transmission is a combination of spin-up and spin-down channels of the P case. In comparison to three layers of $MoS_2$ with Fe(001) electrode, the transmission for three-layer $MoS_2$ with Fe(111) as the electrode, the transmission at zero-bias in the P orientation of electrodes has reduced by order of 10 and is 0.001 and 0.004 for spin-up and spin-down channels, respectively, for the AP orientation of electrodes, both the spin channel has a transmission of 0.002. The reduction in transmission is due to the nearly half-metallic nature of the interface $MoS_2$ layer and the effect of strain at the interface. For Fe(001)/$Mo_xCr_{1-x}S_2$/Fe(001) junctions with doping of magnetic impurity Cr in the middle layer of three, five, and seven-layer $MoS_2$, Cr defect levels are formed below the conduction band of the doped $MoS_2$ layer. The addition of magnetic impurity does not cause significant change in the spin-polarisation of the DOS for the doped $MoS_2$ layer. Higher doping concentration may improve the spin-polarisation of the DOS.

For a thicker junction of five layers and seven layers of $MoS_2$, the middle layer of five-layer $MoS_2$ and the middle layer and three layers of $MoS_2$ in the middle of the seven-layer $MoS_2$ keep a semiconducting band gap and the tunnelling nature of the junction is observed as the interface effect is limited to the first two layers of $MoS_2$ at the interface. The tunnelling spin-current is higher in a 5-layer junction compared to a 7-layer junction, as with a 7-layer thick junction, the bulk behaviour of $MoS_2$ reappears. The 3-layer, 5-layer, and Cr-doped 7-layer devices are stable up to a bias of 0.5V. The 7-layer device is stable up to a bias of 0.7V, and the stability of the junction with respect to bias voltage is related to the band gap of the spacer.

Supplementary file for

# Spin-dependent Transport Studies of $Fe/Mo_xCr_{1-x}S_2/Fe$ Magnetic Tunnel Junction

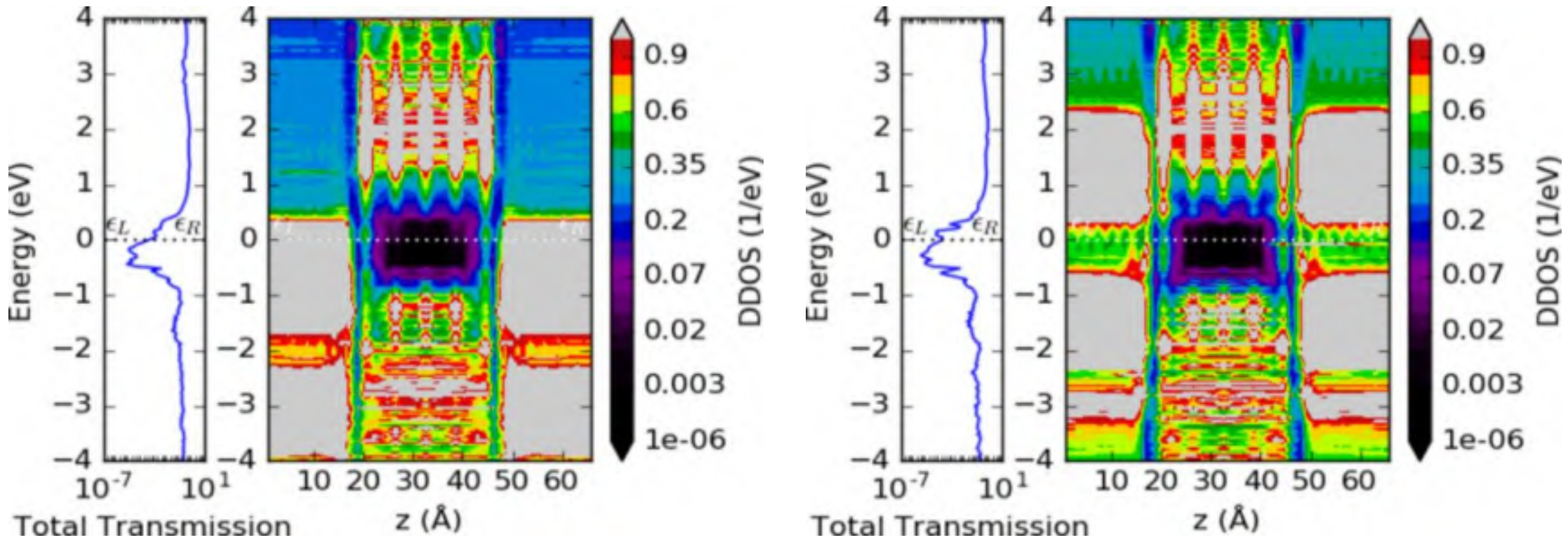

(a) (b

*Figure S1 PLDOS for spin-up (a) and spin-down (b) channel in Fe(001)/$MoS_2$(5-layer)/Fe(001) junction in parallel orientation of electrodes.*

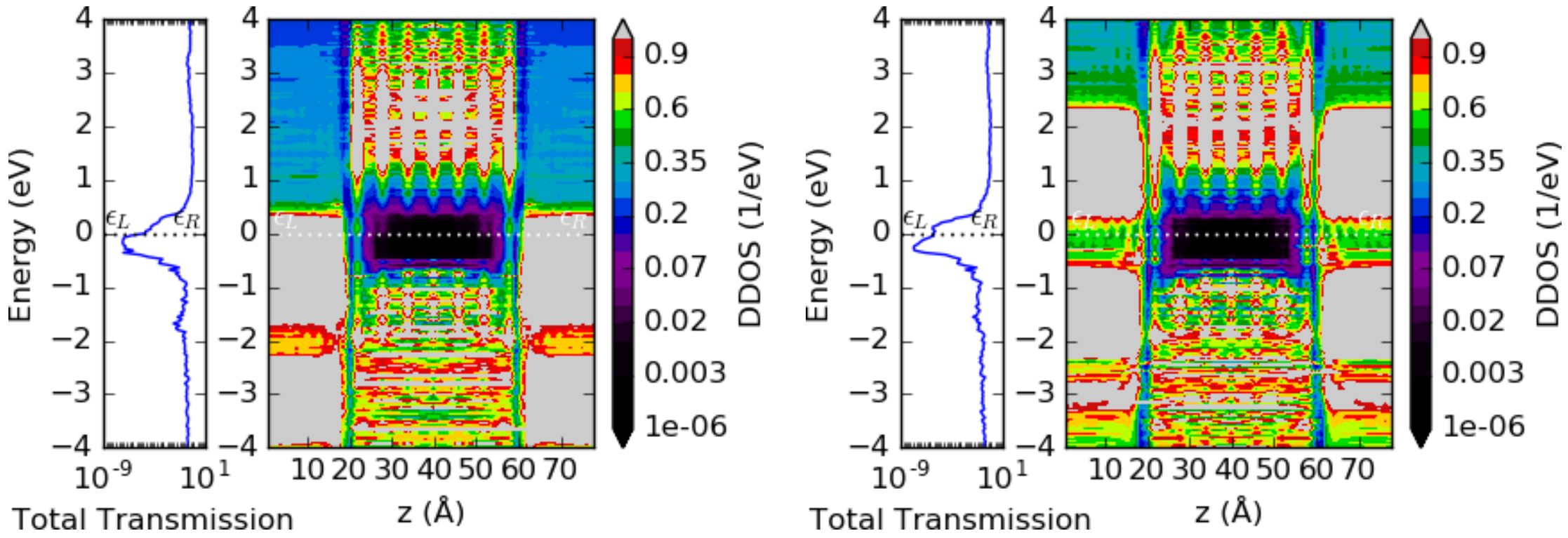

(a) (b)

*Figure S2 PLDOS for spin-up (a) and spin-down (b) channel in Fe(001)/$MoS_2$(7-layer)/Fe(001) junction in parallel orientation of electrodes.*

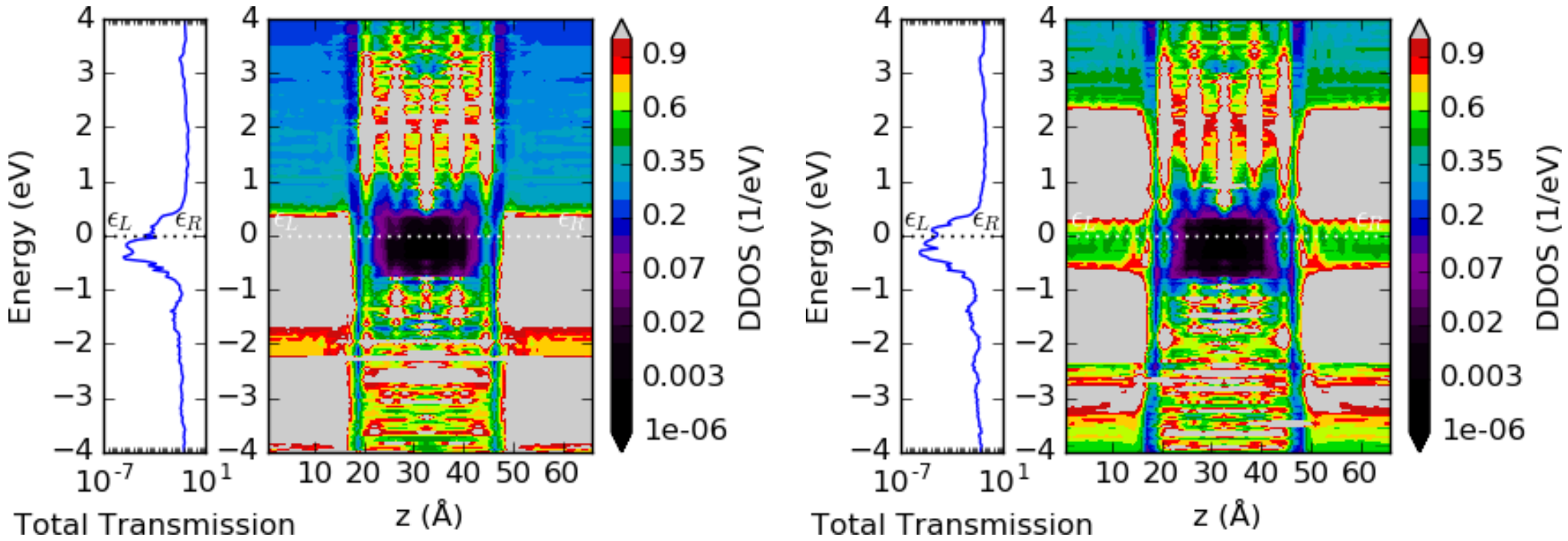

(a) (b)

*Figure S3 PLDOS for spin-up (a) and spin-down channel (b) in Fe(001)/$Mo_xCr_{1-x}S_2$(5-layer)/Fe(001) junction in parallel orientation of electrodes.*

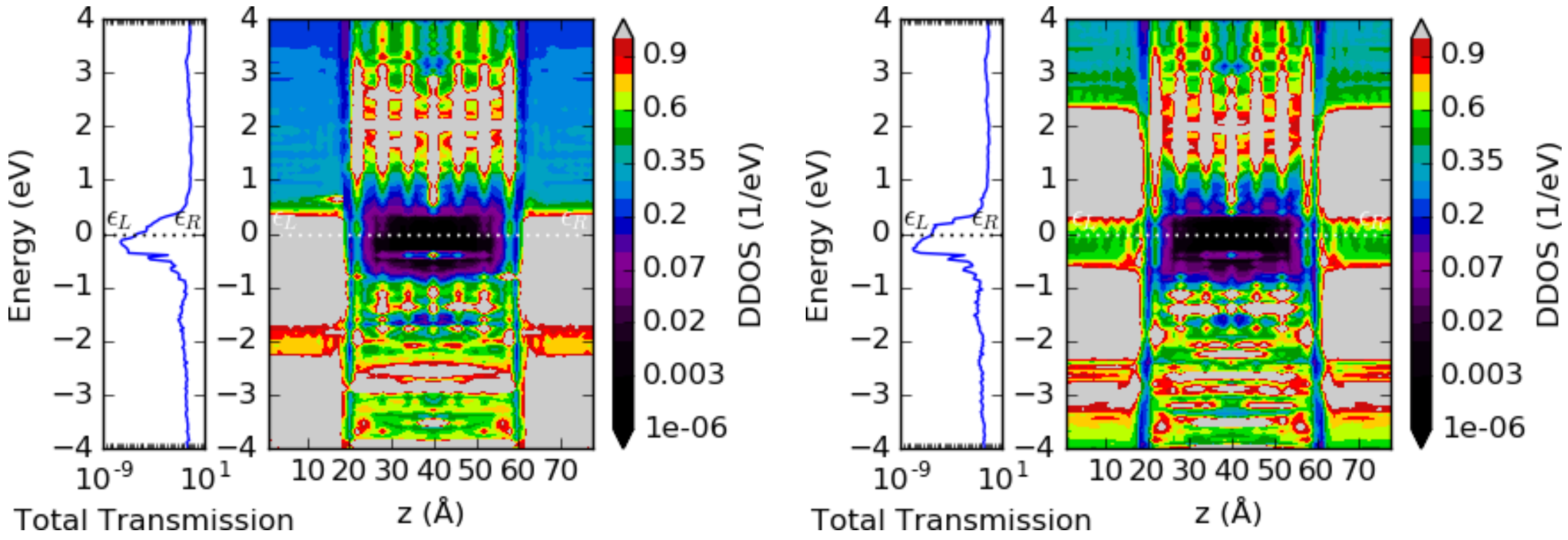

(a) (b)

*Figure S4 PLDOS for spin-up (a) and spin-down (b) channel in Fe(001)/$Mo_xCr_{1-x}S_2$(7-layer)/Fe(001) junction in parallel orientation of electrodes.*

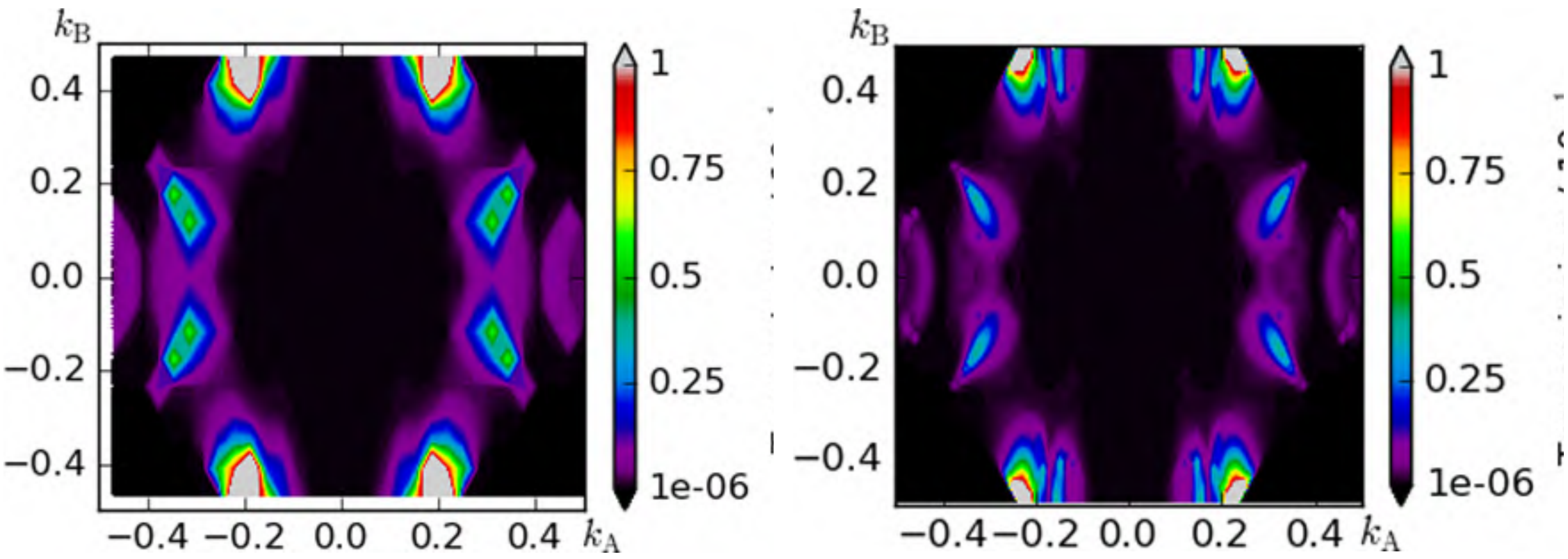

(a) (b)

*Figure S5 K-resolved transmission spectrum in anti-parallel orientation of electrodes for Fe(001)/$Mo_xCr_{1-x}S_2$(3-layer)/Fe(001) junction.*

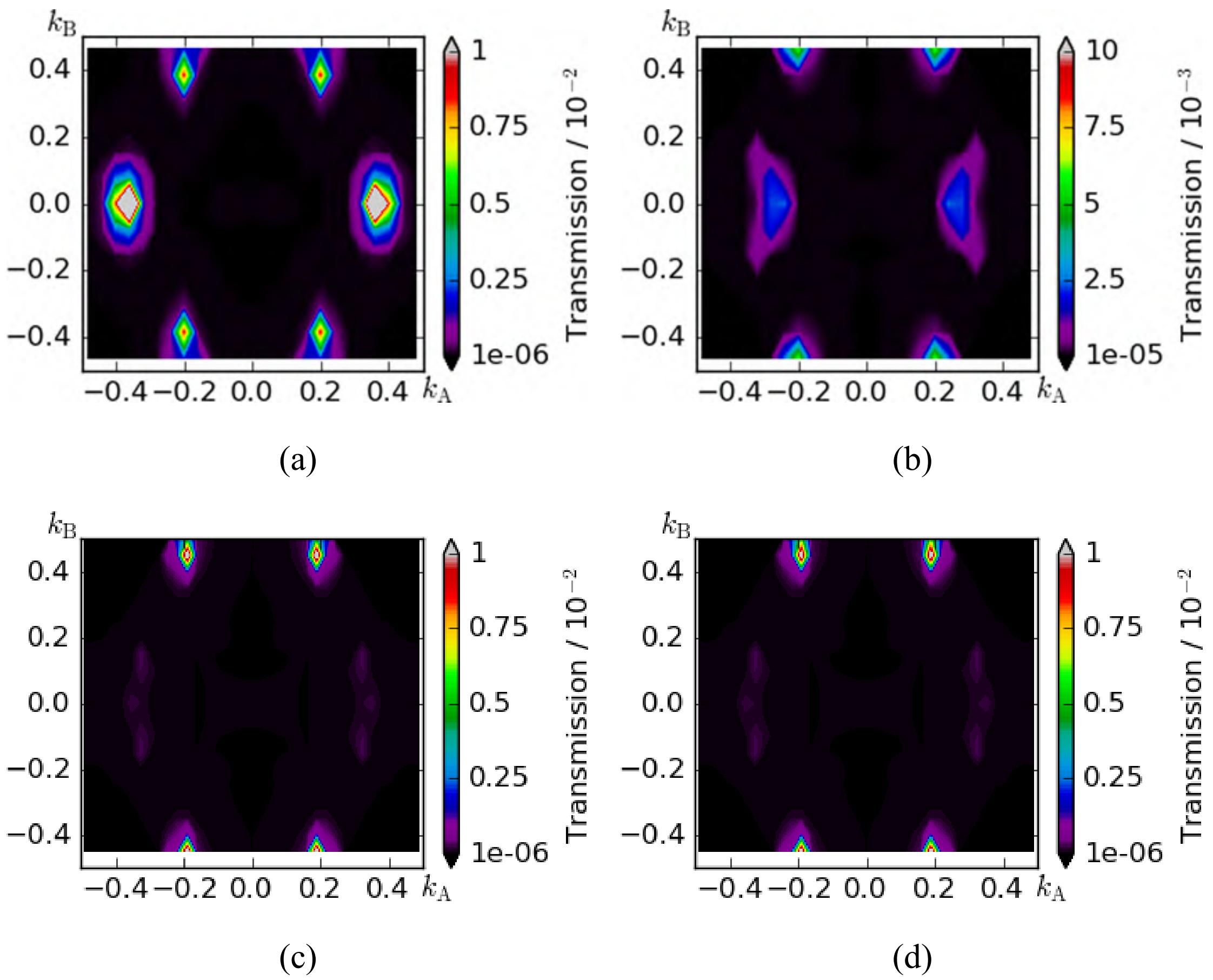

*Figure S6 K-resolved transmission spectrum in spin-up (left panel) and spin-down (right panel) channel for Fe(001)/$MoS_2$(5-layer)/Fe(001) junction in parallel (figure a, b) and anti-parallel (figure c, d) orientation of electrodes.*

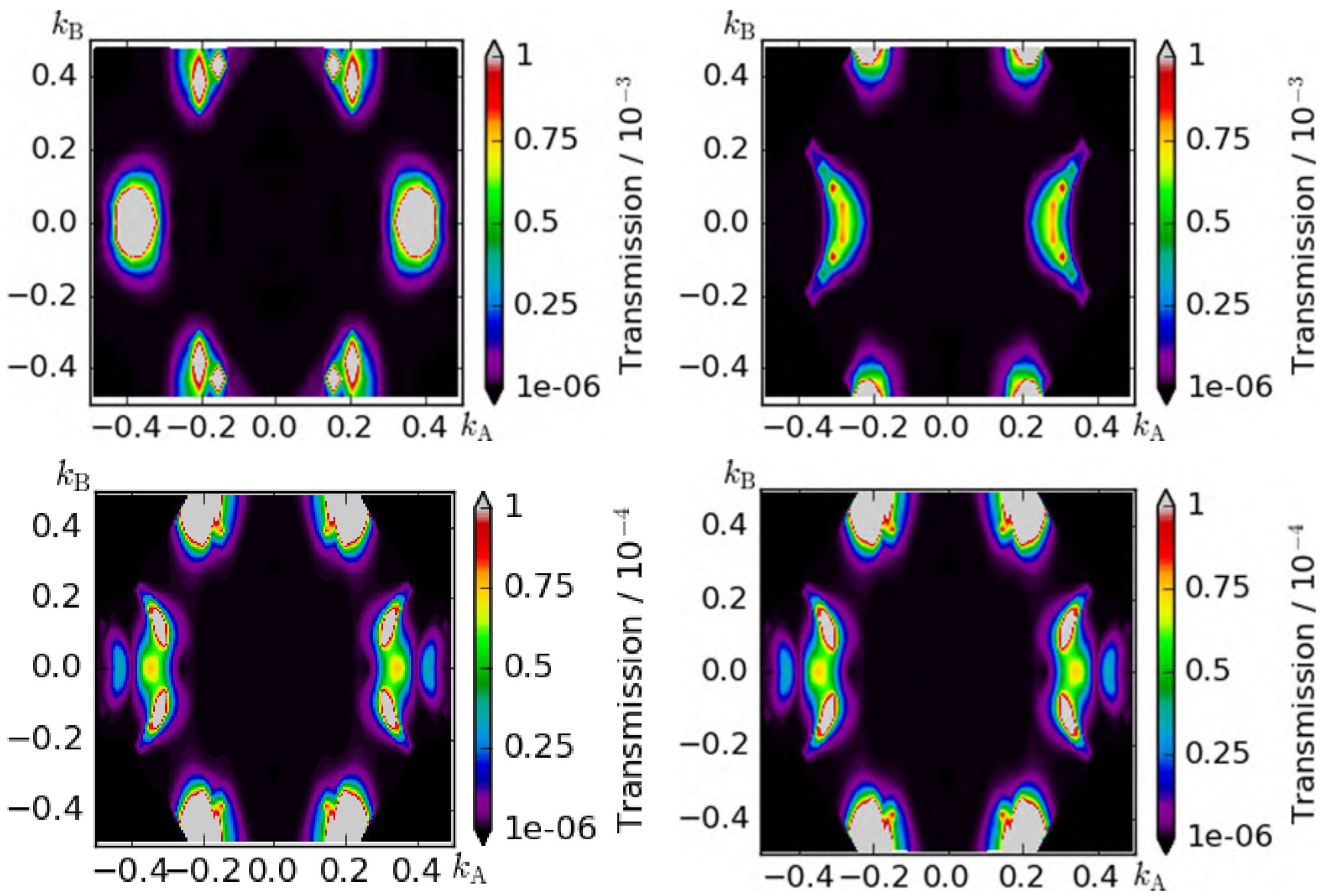

*Figure S7 K-resolved transmission spectrum in spin-up (a) and spin-down (b) channel for Fe(001)/$Mo_xCr_{1-x}S_2$(5-layer)/Fe(001) junction in parallel orientation of electrodes.*

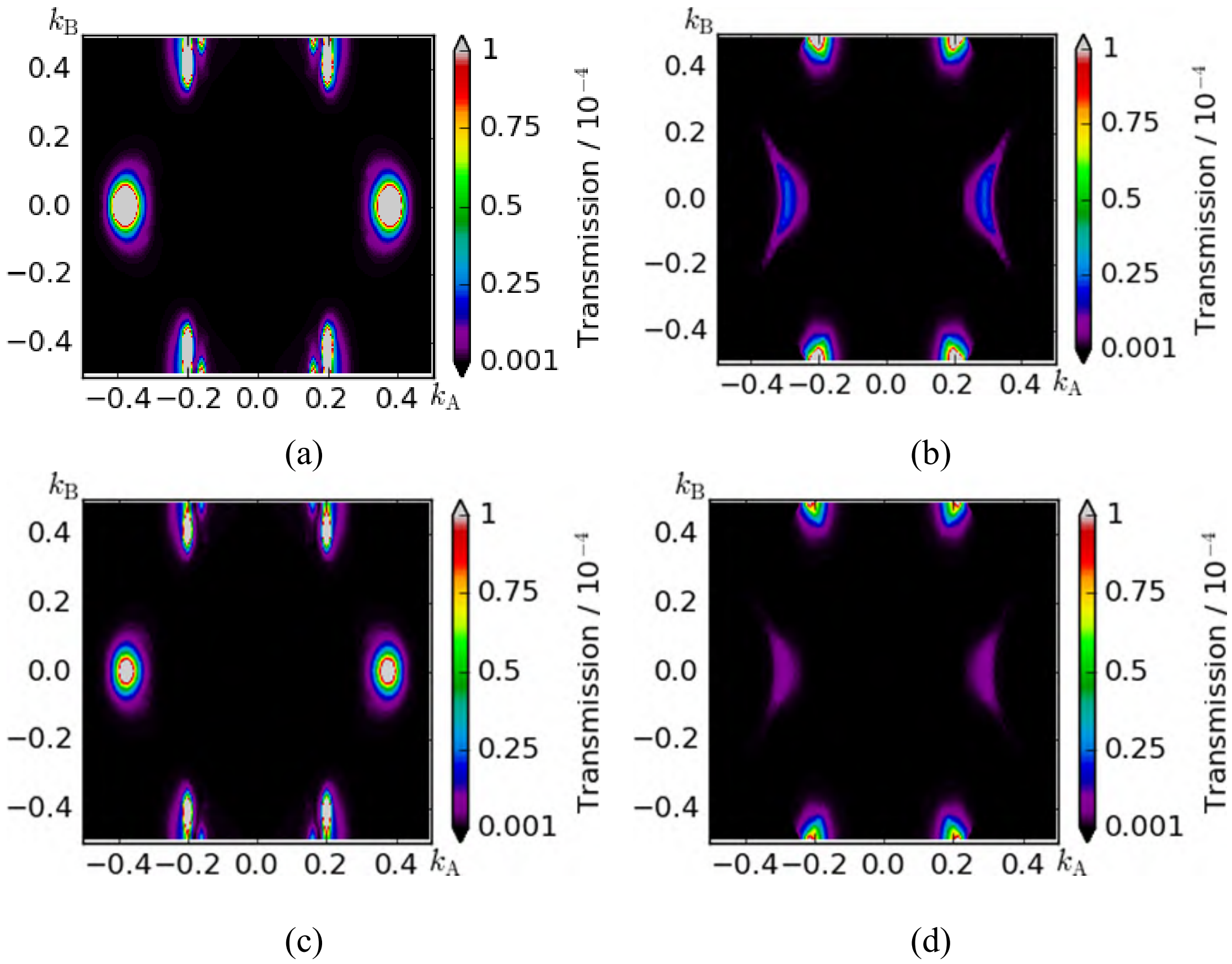

*Figure S8 K-resolved transmission spectrum undoped (upper panel) and Cr-doped (lower panel) Fe(001)/$MoS_2$(7-layer)/Fe(001) junction in the spin-up (a) and spin-down (b) channel in parallel orientation of electrodes.*

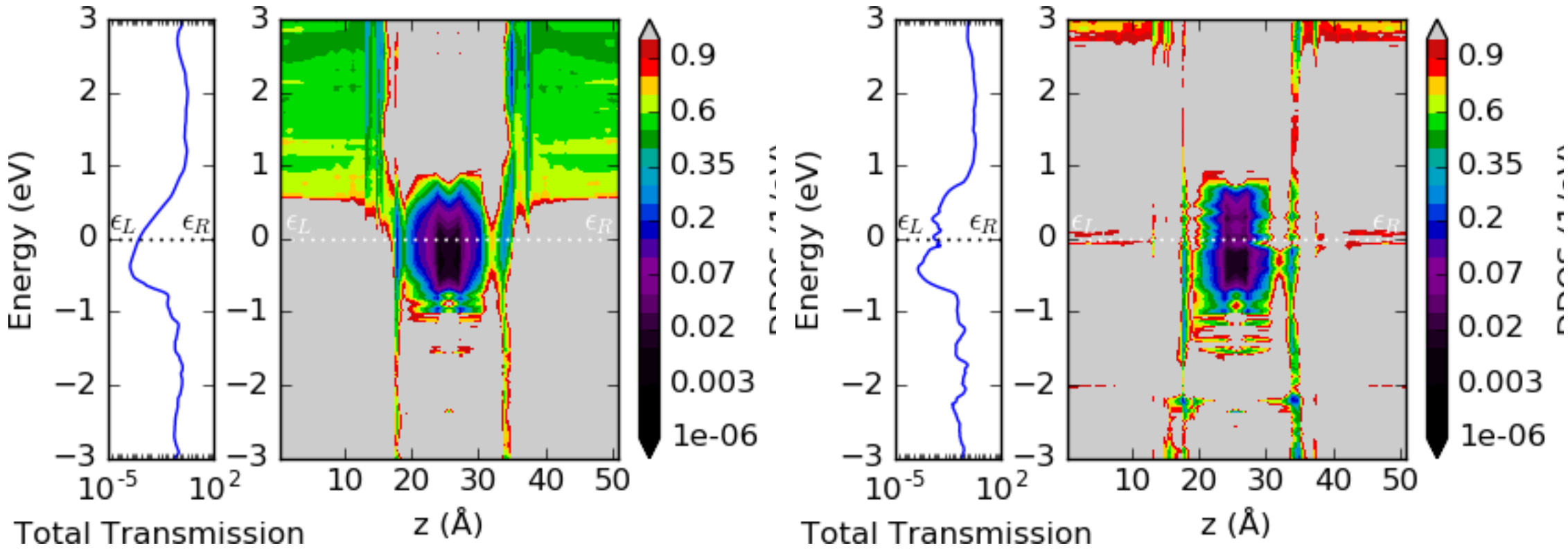

(c) (d)

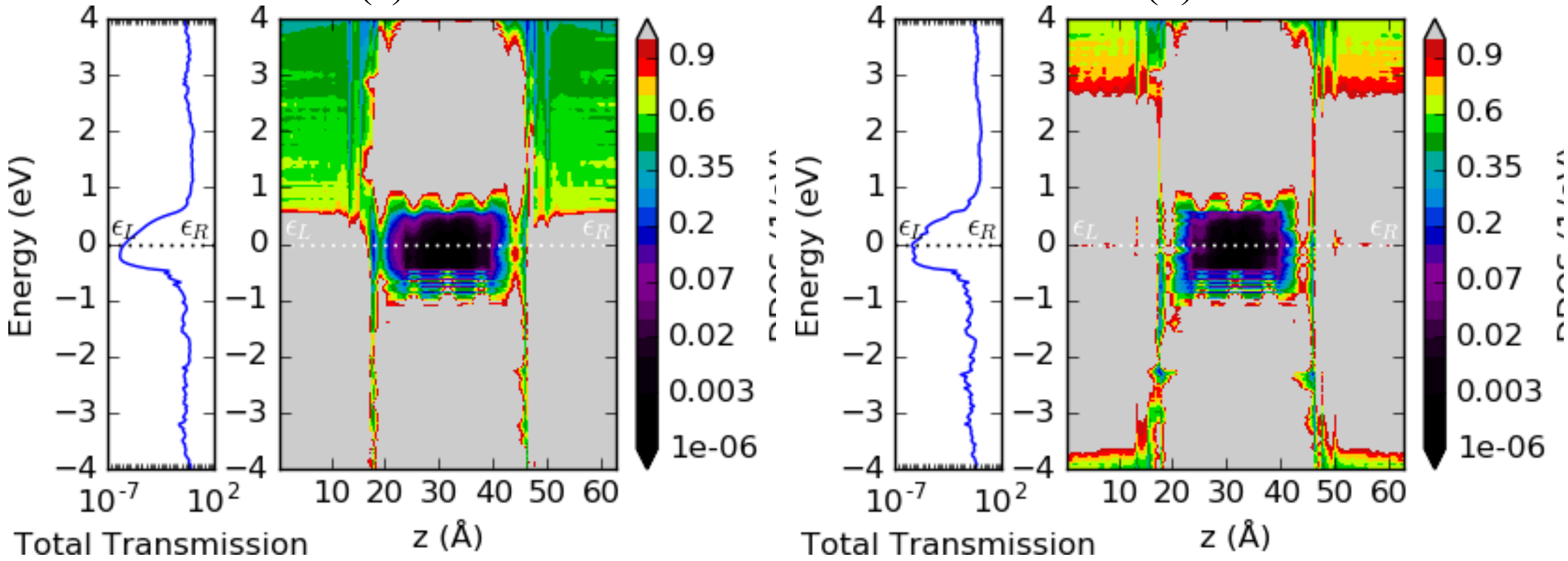

(e) (f)

*Figure S9 PLDOS of one,three, and five layer Fe(111)/$MoS_2$/Fe(111) junction in the parallel orientation of electrodesfor the spin-up (left panel) and spin-down (right panel) channel.*

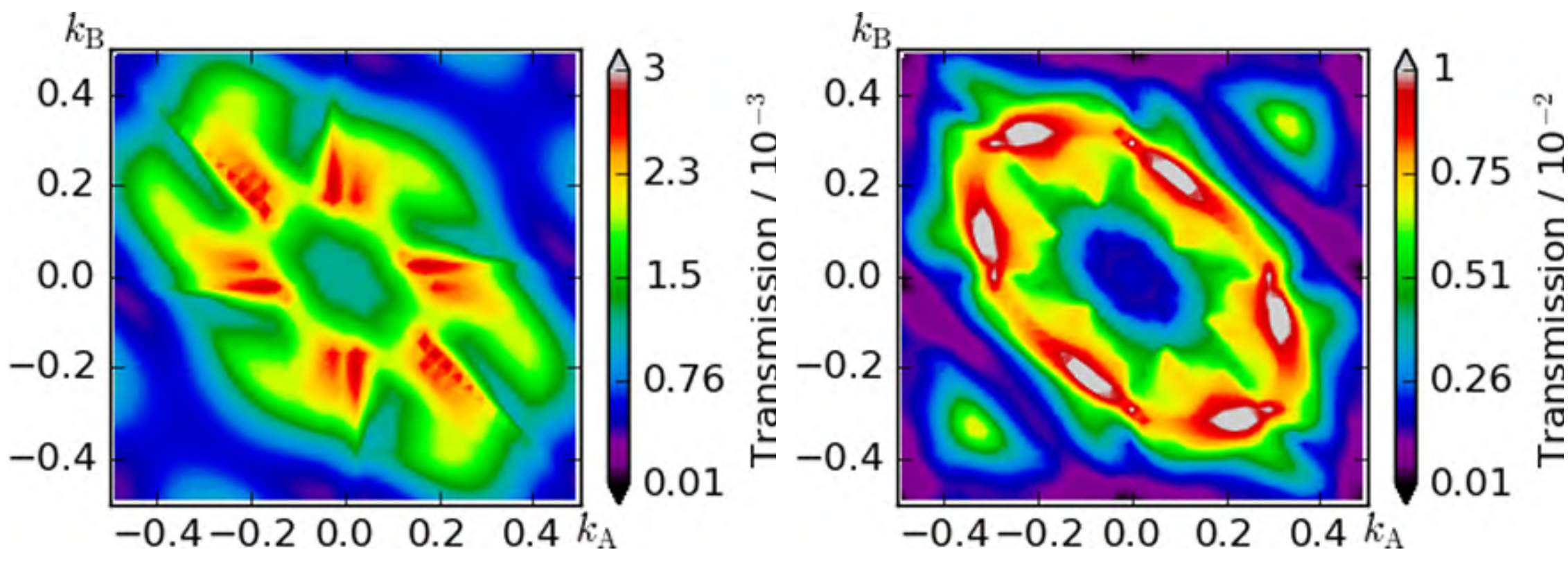

(a) (b)

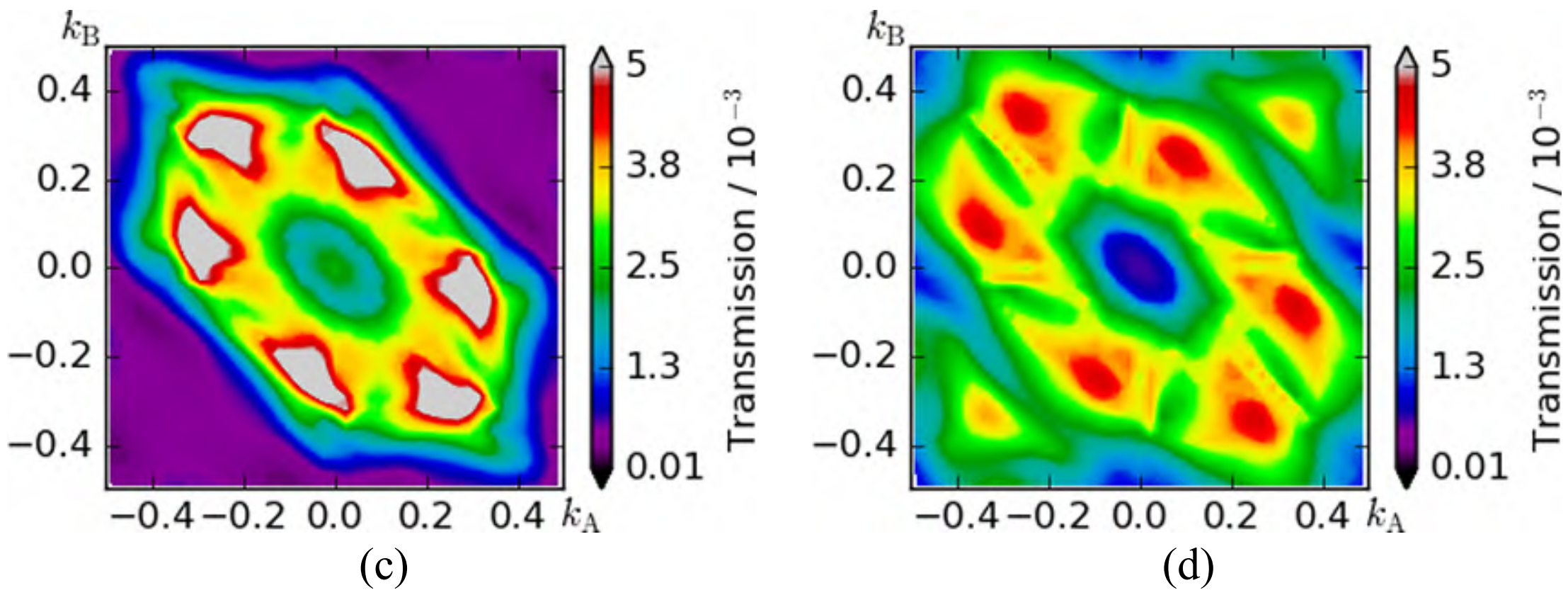

*Figure S10 K-resolved transmission spectrum in spin-up (a) and spin-down (b) channel for Fe(111)/$MoS_2$(3-layer)/Fe(111) junction in parallel orientation of electrodes and in anti-parallel orientation of electrodes spin-up (a) and spin-down(d).*

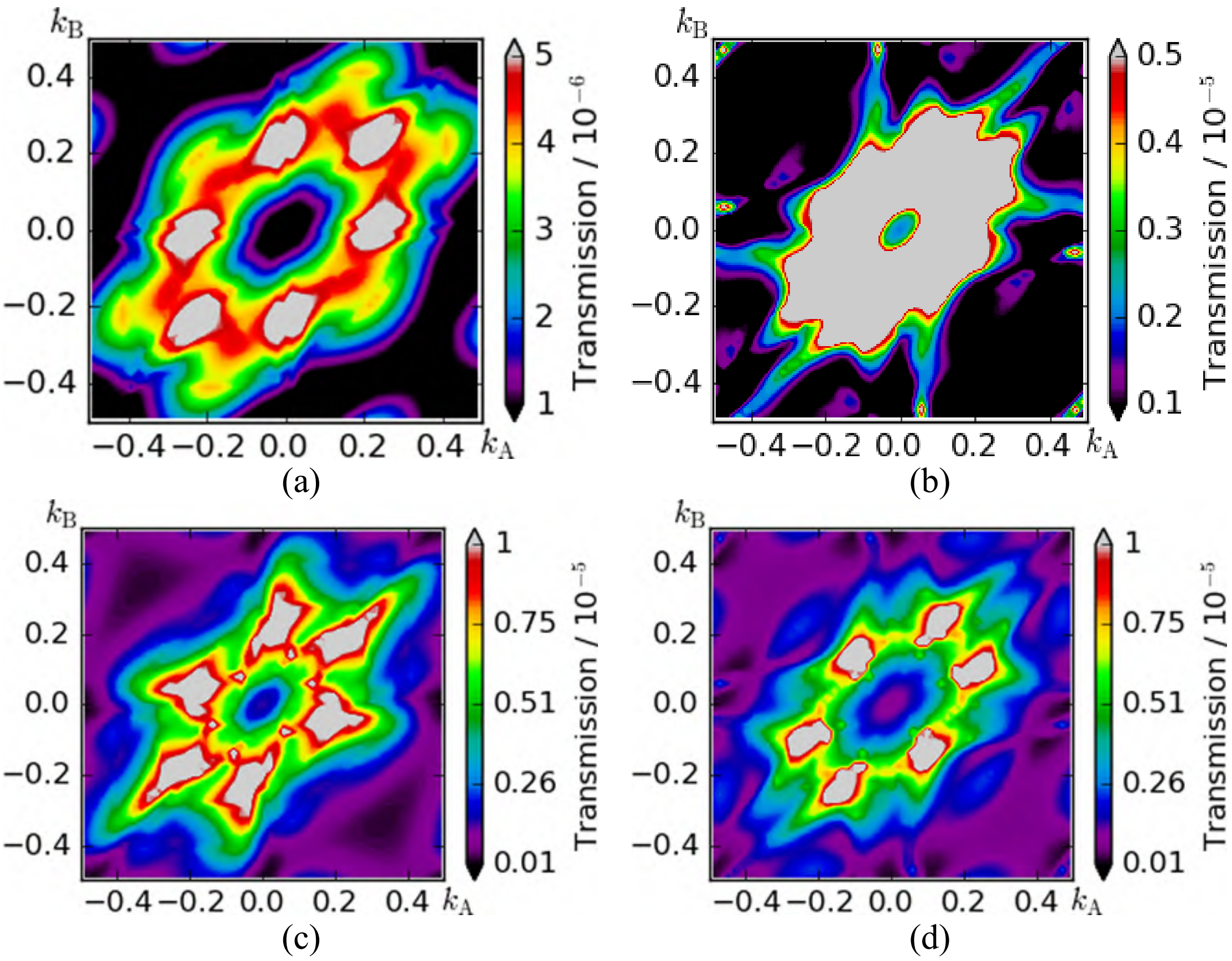

*Figure S11 K-resolved transmission spectrum for spin-up (left panel) and spin-down (right panel) channel of Fe(111)/$MoS_2$(5-layer)/Fe(111) junction in the parallel (a) and (b) and anti-parallel (c) and (d) orientation of electrodes.*

The crystalline configurations ferromagnetic Fe electrode and the spacer $MoS_2$ are first obtained from the bcc unit cell of Fe and the 2H $MoS_2$ having space group P$6_3$/mmc with the help of DFT calculations. The computed electronic structure results compare well with the experimental results. The band gap for bulk $MoS_2$ with LDA approximation gives an indirect band gap of 0.50 eV, and direct band gap of 1.757 eV. The computed band gap for monolayer $MoS_2$ has a direct band gap of 1.88 eV well in agreement with the experimental values. The relaxed lattice of bcc Fe has lattice parameter a = 2.77 Å, which is within the 10% error limit of LDA.

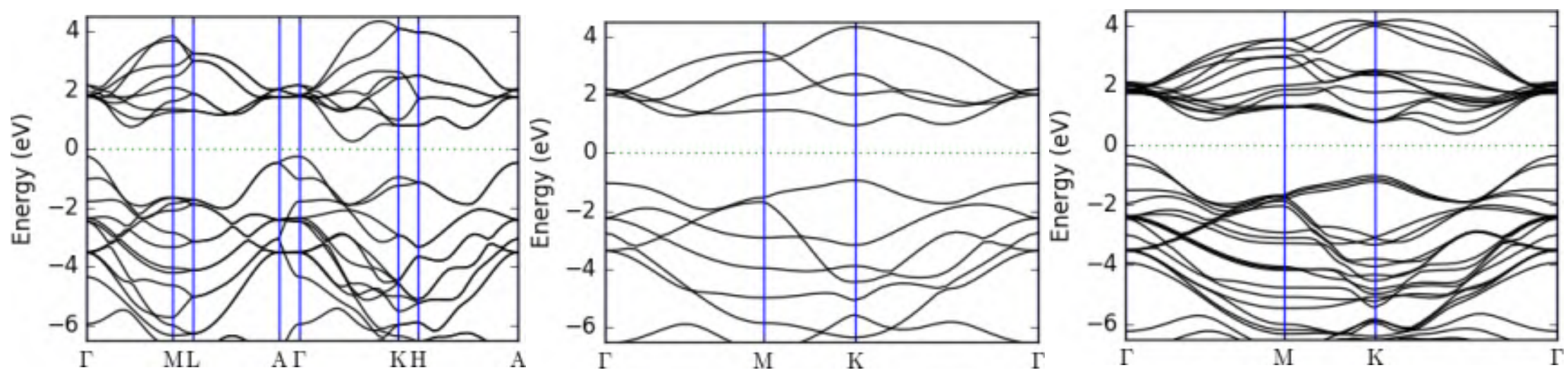

*Figure S12 Bandstructure for (a) Bulk $MoS_2$, (b) Monolayer $MoS_2$ and (c) Three-layer $MoS_2$.*